\documentclass[journal,twoside,web]{ieeecolor}
\usepackage{generic}
\usepackage{cite}
\usepackage{amsmath,amssymb,amsfonts}
\usepackage{algorithm,algorithmic}
\usepackage{graphicx}
\usepackage{hyperref}
\hypersetup{hidelinks=true}
\usepackage{textcomp}
\usepackage{booktabs}
\let\labelindent\relax
\usepackage{enumitem}
\usepackage[caption=false,font=footnotesize]{subfig}

\def\BibTeX{{\rm B\kern-.05em{\sc i\kern-.025em b}\kern-.08em
    T\kern-.1667em\lower.7ex\hbox{E}\kern-.125emX}}

\begin{document}

\title{Dual-Timescale Hebbian Accumulators for Online Spiking Neural Network Decoding in Intracortical Brain Machine Interfaces}

\author{Sriram V.C. Nallani
    and Sahil Shah
\thanks{Manuscript received XXXX XX, 2026; revised XXXX XX, 2026.}
\thanks{S.V.C. Nallani and S. Shah are with the University of Maryland,
    College Park, MD 20742 USA.}
\thanks{Corresponding author: Sriram V.C. Nallani.}}

\maketitle

\begin{abstract}
Intracortical brain-machine interfaces require decoders that adapt continuously to neural signal instability while operating within strict memory budgets. We introduce a dual-timescale Hebbian accumulator learning rule for spiking neural networks that enables per-timestep online supervised updates with training memory constant in sequence length, avoiding backpropagation through time. The rule combines synapse-specific fast and slow eligibility traces, error-modulated three-factor updates, and integer-friendly RMS homeostasis, operating without adaptive gradient optimizers (Adam, RMSProp) or replay buffers. On two primate intracortical datasets, the method achieves Pearson correlations of $R \geq 0.81$ on MC~Maze and $R \geq 0.63$ on Zenodo~Indy, with 63--86\% measured memory reduction versus BPTT at sequence length $T = 1000$. Closed-loop simulations demonstrate online adaptation to neural disruptions and learning from scratch without offline calibration.
\end{abstract}


\section{Introduction}
\label{sec:intro}

\IEEEPARstart{I}{ntracortical} brain--computer interfaces (BMIs) translate neural activity recorded from Utah-array microelectrode implants into control signals, bypassing conventional neuromuscular pathways~\cite{Pandarinath2017,Collinger2013,Bouton2016}. These systems face persistent signal instability arising from electrode encapsulation, micromotion, and neural plasticity, which degrades decoder performance over time and necessitates continuous online adaptation~\cite{Dangi2014,Ganguly2009,Sussillo2016,Woeppel2021,Maynard1997}.

Current approaches occupy two extremes. Recurrent artificial neural networks (LSTM, GRU) achieve high offline decoding accuracy but require backpropagation through time (BPTT) with $O(T)$ activation memory and batch retraining, limiting their suitability for long-running, memory-constrained implantable systems~\cite{Huang2021,Premchand2020}. Kalman filters and delta-rule updates adapt online with constant memory but are limited to linear or shallow mappings that cannot capture the nonlinear temporal structure of intracortical signals~\cite{Wu2004,Jiang2018}. No existing method provides online per-timestep updates with nonlinear temporal modeling and training memory constant in sequence length.

This work introduces a dual-timescale Hebbian accumulator learning rule for spiking neural networks (SNNs) that addresses this gap. The contributions are:
\begin{enumerate}[nosep]
  \item A learning rule that replaces BPTT with synapse-specific dual-timescale eligibility traces acting as Hebbian accumulators, requiring only fixed-size buffers independent of sequence length.
  \item An integer-friendly implementation using bitshift RMS homeostasis, operating without adaptive gradient optimizers or replay buffers, suitable for resource-constrained deployment.
  \item Evaluation on two primate intracortical datasets (MC~Maze, Zenodo~Indy) and closed-loop simulations demonstrating online adaptation to neural disruptions and learning from scratch without offline calibration.
\end{enumerate}

\section{Related Work}
\label{sec:related}

\textbf{Adaptive BMI decoders.}
Traditional BMI decoders range from Kalman filters with closed-form parameter estimation~\cite{Wu2004} to deep recurrent networks trained with BPTT~\cite{Huang2021,Premchand2020}. Co-adaptive systems address decoder--user interaction but generally rely on batch updates rather than continuous per-timestep online learning~\cite{Carmena2013,Dangi2014,Orsborn2014,Bhaduri2018}. From a hardware perspective, neuromorphic implementations favor local, on-chip learning rules~\cite{Basu2022SNNICReview}.

\textbf{Temporally local SNN learning rules.}
Three-factor extensions of STDP incorporate modulatory signals: reward-modulated STDP uses dopamine-like signals~\cite{Florian2007,Fremaux2016,Izhikevich2007}, while eligibility trace methods such as SuperSpike and e-prop use broadcast error signals with local traces~\cite{Zenke2018,Bellec2020}. Recent rules including OTTT, S-TLLR, SLTT, and TESS~\cite{Xiao2022OTTT,ApolinarioRoy2025STLLR,Meng2023SLTT,ApolinarioRoyFrenkel2025TESS} explore temporally and spatially local updates for large-scale computer vision benchmarks. These methods predominantly use neuron-level $O(N)$ trace factorization and floating-point optimizers tuned on homogeneous image inputs. The present method uses synapse-specific dual-timescale Hebbian accumulators with RMS homeostasis, targeting heterogeneous intracortical recordings at Utah-array scales.

\section{Method}
\label{sec:method}

\subsection{Problem Setup}
\label{subsec:problem}

We observe spike count vectors $\mathbf{x}_t \in \mathbb{R}^{N}$ and predict 2D velocity $\mathbf{y}_t \in \mathbb{R}^2$ at time bins $t = 1, \ldots, T$. Hidden layer $k$ has membrane potentials $\mathbf{u}^{(k)}_t$ and spikes $\mathbf{s}^{(k)}_t$, with the first layer maintaining recurrent state $\mathbf{s}^{(1)}_{t-1}$. The final layer's membrane potentials directly represent velocity predictions: $\hat{\mathbf{y}}_t = \mathbf{u}^{(3)}_t$. Training minimizes per-timestep squared error $\mathcal{L}_t = \|\hat{\mathbf{y}}_t - \mathbf{y}_t\|_2^2$ using a local three-factor rule with dual eligibility traces. No unrolled computational graph or replay buffer is required.

\subsection{Architecture}
\label{subsec:architecture}

\begin{figure}[!t]
    \centerline{\includegraphics[width=0.85\columnwidth]{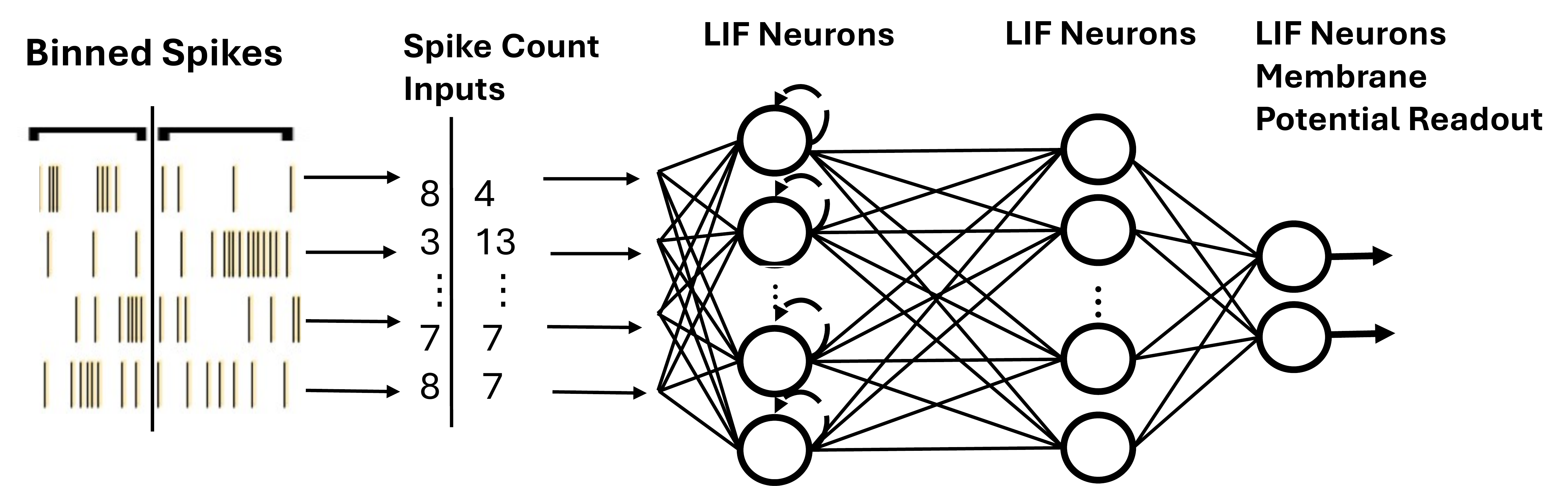}}
    \caption{Three-layer LIF architecture with recurrent connections in the first hidden layer and 2D velocity output.}
    \label{fig:architecture}
\end{figure}

The network comprises three layers of Leaky Integrate-and-Fire (LIF) neurons (Fig.~\ref{fig:architecture}) with recurrent connections in the first hidden layer. Background on SNN computation is provided in Appendix~\ref{app:overviewSNN}. Each neuron's membrane potential integrates synaptic currents and decays with factor $\beta$:
\[
u_i(t+1) = \beta\,u_i(t) + I_i(t) - s_i(t).
\]
Because the Heaviside spike function $s_i = \Theta(u_i - \theta)$ has zero derivative almost everywhere, we define a local neuronal sensitivity using the fast sigmoid surrogate~\cite{Zenke2018,Neftci2019}:
\begin{equation}
d_{\text{LIF},i}(t) = \frac{1}{\big(1 + k\,|u_i(t) - \theta|\big)^{2}}, \quad k=25, \tag{S}
\end{equation}
where $\theta{=}1.0$ is the firing threshold. This is not part of a backward pass: no computational graph is constructed or unrolled. Instead, $d_{\text{LIF}}$ is evaluated locally at each neuron's current membrane potential and used directly as the postsynaptic sensitivity factor in the three-factor Hebbian update (Eq.~1), concentrating plasticity on neurons near threshold. The network processes raw spike counts directly; no rate normalization is applied to SNN inputs.

\subsection{Learning Rule}
\label{subsec:learning_rule}

For layer $\ell$, let $W^{(\ell)}\!\in\!\mathbb{R}^{n_{\mathrm{post}}\times n_{\mathrm{pre}}}$ be the weight matrix, $\mathrm{pre}^{(\ell)}_t$ the presynaptic activity, $\mathbf{e}^{(\ell)}_t$ the layer-local error drive, and $d^{(\ell)}_t$ the postsynaptic sensitivity (LIF surrogate gradient). We set $\mathrm{pre}^{(1)}_t=\mathbf{x}_t$, $\mathrm{pre}^{(2)}_t=\mathbf{s}^{(1)}_t$, $\mathrm{pre}^{(3)}_t=\mathbf{s}^{(2)}_t$, and $\mathrm{pre}^{(\mathrm{rec})}_t=\mathbf{s}^{(1)}_{t-1}$ for the recurrent path.

\textbf{Error computation.}
Hidden-layer error signals are computed via layer-wise error backpropagation using the transpose of each layer's weight matrix, with RMS normalization applied at every layer to maintain well-scaled gradients:
\begin{align*}
\tilde{\mathbf{e}}^{(3)}_t &:= \mathcal{R}\!\big(\mathbf{y}_t-\hat{\mathbf{y}}_t\big),\\
\mathbf{e}^{(2)}_t &:= \mathcal{R}\!\Big(\big(W^{(3)}\big)^{\!\top}\tilde{\mathbf{e}}^{(3)}_t\Big),\\
\mathbf{e}^{(1)}_t &:= \mathcal{R}\!\Big(\big(W^{(2)}\big)^{\!\top}\mathbf{e}^{(2)}_t\Big).
\end{align*}
Each $\mathcal{R}(\cdot)$ uses a per-unit exponential moving average of squared magnitudes to track the running RMS, ensuring that error signals remain well-conditioned across layers and across sessions with varying noise characteristics.  All updates depend only on current-timestep quantities and fixed-size eligibility traces. Recent work has demonstrated that this weight-transpose operation can be implemented fully on-chip on neuromorphic hardware without an external computer in the loop~\cite{Renner2024}.

\textbf{Three-factor Hebbian updates.}
Plasticity at each synapse is driven by the coincidence of presynaptic activity, postsynaptic sensitivity, and error:
\begin{equation}
\Delta W^{(\ell)}_{\mathrm{hebb}}(t) = \big(\tilde{\mathbf{e}}^{(\ell)}_t \odot d^{(\ell)}_t\big)\,\big(\mathrm{pre}^{(\ell)}_t\big)^{\!\top}. \tag{1}
\end{equation}
The same rule updates $W^{(\mathrm{rec})}$ with $\mathrm{pre}^{(\mathrm{rec})}_t$. Rather than applying these updates directly, each synapse accumulates them in dual eligibility traces with different decay timescales:
\begin{align}
E^{(\ell)}_{\mathrm{fast}}(t) &= \lambda_{\mathrm{fast}}\,E^{(\ell)}_{\mathrm{fast}}(t-1) + \Delta W^{(\ell)}_{\mathrm{hebb}}(t), \tag{2}\\
E^{(\ell)}_{\mathrm{slow}}(t) &= \lambda_{\mathrm{slow}}\,E^{(\ell)}_{\mathrm{slow}}(t-1) + \Delta W^{(\ell)}_{\mathrm{hebb}}(t), \tag{3}
\end{align}
where $\lambda_{\mathrm{fast}}=\exp(-\Delta t/\tau_{\mathrm{fast}})$ and $\lambda_{\mathrm{slow}}=\exp(-\Delta t/\tau_{\mathrm{slow}})$. In True Online mode, time constants are scaled for faster adaptation: $\tau_{\mathrm{fast}}$ is halved and $\tau_{\mathrm{slow}}$ is multiplied by 0.8. These are combined as:
\begin{equation}
E^{(\ell)}_{\mathrm{comb}}(t) = \alpha_{\mathrm{mix}}\,E^{(\ell)}_{\mathrm{fast}}(t) + \big(1-\alpha_{\mathrm{mix}}\big)\,E^{(\ell)}_{\mathrm{slow}}(t). \tag{4}
\end{equation}

The fast trace emphasizes recent three-factor updates; the slow trace averages over a longer history. The empirical advantage of synapse-specific over neuron-level factorized traces is quantified in Section~\ref{subsec:ablations}.

\textbf{Dual-timescale weight updates.}
The combined trace drives fast updates applied at each timestep:
\[
W^{(\ell)} \leftarrow W^{(\ell)} + \eta_{\mathrm{fast}}\,E^{(\ell)}_{\mathrm{comb}}(t).
\]
A momentum-smoothed accumulator builds evidence for stable consolidation:
\[
G^{(\ell)}(t) = \mu\,G^{(\ell)}(t-1) + (1-\mu)\,E^{(\ell)}_{\mathrm{comb}}(t).
\]
Every $K$ timesteps, this drives slow updates:
\[
W^{(\ell)} \leftarrow W^{(\ell)} + \eta_{\mathrm{slow}}\,\mathcal{R}\!\big(\bar G^{(\ell)}_K(t)\big),
\]
where $\mathcal{R}(\cdot)$ provides RMS normalization and $\bar G^{(\ell)}_K(t)$ is the $K$-step average. The fast pathway enables rapid adaptation; the slow pathway preserves stable structure~\cite{Pfister2006,Benna2016}. The complete per-timestep update is given in Algorithm~\ref{alg:online_update}.

\begin{algorithm}[!t]
\caption{Per-Timestep Online Update}\label{alg:online_update}
\begin{algorithmic}
\STATE \textbf{Input:} Spikes $\mathbf{x}_t$, target velocity $\mathbf{y}_t$
\STATE \textbf{State:} Weights $W$, potentials $\mathbf{u}(t)$, traces $E^{\text{fast}}, E^{\text{slow}}$
\STATE
\STATE 1. \textbf{Forward:} $\mathbf{u}(t) \leftarrow \beta\,\mathbf{u}(t{-}1) + W\mathbf{x}(t)$; get $\hat{\mathbf{y}}_t, d_{\text{LIF}}$
\STATE 2. \textbf{Error:} $\mathbf{e}(t) \leftarrow \mathbf{y}(t) - \hat{\mathbf{y}}(t)$
\STATE 3. \textbf{RMS norm:} Update EMAs; $\tilde{\mathbf{e}}(t) \leftarrow \mathcal{R}(\mathbf{e}(t))$
\STATE 4. \textbf{Three-factor:} $\Delta W^{\text{hebb}} \leftarrow (\tilde{\mathbf{e}} \odot d_{\text{LIF}})\,(\mathrm{pre}_t)^T$
\STATE 5. \textbf{Traces:} $E^{\text{fast}} \leftarrow \lambda_f\,E^{\text{fast}} + \Delta W^{\text{hebb}}$
\STATE \hspace{2.85em} $E^{\text{slow}} \leftarrow \lambda_s\,E^{\text{slow}} + \Delta W^{\text{hebb}}$
\STATE 6. \textbf{Combine:} $E^{\text{comb}} \leftarrow \alpha_{\text{mix}}\,E^{\text{fast}} + (1{-}\alpha_{\text{mix}})\,E^{\text{slow}}$
\STATE 7. \textbf{Fast update:} $W \leftarrow W + \eta_{\text{fast}}\,E^{\text{comb}}$
\STATE 8. \textbf{Accumulate:} $G \leftarrow \mu\,G + (1{-}\mu)\,E^{\text{comb}}$
\STATE 9. \textbf{if} $t \bmod K = 0$: $W \leftarrow W + \eta_{\text{slow}} \cdot \mathcal{R}(\bar{G}_K)$; reset $G, E^{\text{fast}}, E^{\text{slow}} \leftarrow 0$
\STATE
\STATE \textbf{Memory:} $O(P)$ static, constant in $T$
\STATE \textbf{Compute:} $O(P)$ per timestep, no backprop graph
\end{algorithmic}
\end{algorithm}

\subsection{Stability Mechanisms}
\label{subsec:stability}

Two mechanisms maintain stable operation during continuous online learning.

\textbf{RMS-based normalization.}
The normalization operator $\mathcal{R}(\mathbf{v}) = \mathbf{v} / (\sqrt{\mathrm{mean}(\mathbf{v}^2)} + \epsilon)$ is applied to error signals and pre-synaptic spike activities, using exponential moving averages of squared magnitudes to track the running RMS. A 256-entry lookup table maps RMS values to inverse square roots, enabling hardware-efficient normalization without floating-point division~\cite{Davies2018,Hassan2024}. This keeps signal magnitudes well-scaled across varying neural activity levels without adaptive gradient optimizers.

\textbf{Per-neuron weight projection.}
After each timestep, the L2 norm of each neuron's incoming weight vector is checked against a threshold $c_\ell = 6.0$. Weights exceeding this threshold are rescaled using power-of-two division, implementable as bitshift operations on neuromorphic hardware.

\subsection{Memory Analysis}
\label{subsec:memory}

\begin{table}[!t]
  \centering
  \caption{Peak Training Memory (MiB) at $T{=}120$ Timesteps}
  \label{tab:memory}
  \resizebox{\columnwidth}{!}{%
  \begin{tabular}{lcccc}
    \toprule
    \textbf{Component} & \textbf{Online} & \textbf{BPTT} & \textbf{Online} & \textbf{BPTT} \\
     & \textbf{96-256-128-2} & \textbf{96-256-128-2} & \textbf{96-1024-512-2} & \textbf{96-1024-512-2} \\
    \midrule
    Parameters $W{+}b$                & 0.47 & 0.47 & 6.38 & 6.38 \\
    Eligibility traces                 & 0.94 & --   & 12.76 & --   \\
    Momentum \& RMS                    & 0.47 & --   & 6.38 & --   \\
    Gradients                          & --   & 0.47 & --   & 6.38 \\
    Optimizer state (Adam)             & --   & 0.94 & --   & 12.77 \\
    \midrule
    Static total                       & 1.88 & 1.88 & 25.53 & 25.54 \\
    Dynamic activations                & 0.00 & 1.39 & 0.00 & 5.26 \\
    \midrule
    \textbf{Peak total}                & \textbf{1.88} & \textbf{3.27} & \textbf{25.53} & \textbf{30.80} \\
    \bottomrule
  \end{tabular}}
\end{table}

The online update stores parameters plus three weight-sized buffers per trained matrix (fast trace, slow trace, momentum/RMS accumulator), yielding static memory on the order of $4P$ where $P$ is the parameter count. This does not grow with sequence length $T$. BPTT additionally stores $O(T)$ activations for gradient computation. Table~\ref{tab:memory} shows measured peak memory at $T{=}120$. Both approaches have $O(P)$ spatial complexity and $O(TP)$ total compute, but removal of the $O(T)$ activation term leads to measured reductions of 63--86\% at $T{=}1000$, with progressively larger savings for longer sequences (see Appendix~\ref{app:memory_scaling}).

\section{Experiments}
\label{sec:experiments}

\subsection{Datasets and Baselines}
\label{subsec:datasets}

\textbf{MC~Maze}~\cite{Churchland2022}: Intracortical multi-unit recordings from a macaque performing 2D maze navigation. 182 input channels (M1 + PMd). Spike counts binned at 100\,ms with 10\,ms stride and 80\,ms kinematic lag. Trial-wise splits (70/15/15\%).

\textbf{Zenodo~Indy}~\cite{ODoherty2017Zenodo}: Self-paced center-out reaching, 96 channels (M1). 50\,ms non-overlapping bins at zero lag. Chronological splits (70/15/15\%).

\textbf{Baselines.} Kalman Filter (KF), LSTM~\cite{Premchand2020}, GRU, MLP, and a BPTT-trained SNN sharing the same architecture as the Online SNN. The BPTT-SNN is trained with Adam ($\mathrm{lr}{=}10^{-3}$) using truncated BPTT over 10-timestep sequences for up to 50 epochs with early stopping (patience 10). GRU and MLP use two 256-unit hidden layers with Adam. SNN variants consume raw spike counts; all other baselines consume z-score-normalized firing rates. The Online SNN is evaluated in two modes: \emph{Batched Online} (10-timestep sequences with 50\% overlap for offline comparison) and \emph{True Online} (single-timestep updates simulating continuous operation). SNN architecture is 182-1024-512-2 for MC~Maze and 96-256-128-2 for Zenodo~Indy.

\textbf{Metrics.} Pearson correlation $R$ between predicted and true velocity components, reported as mean $\pm$ SEM ($n{=}10$). Closed-loop tasks report time-to-target. Statistical comparisons use Welch's $t$-test with thresholds: {*}\,$p < 0.05$, {**}\,$p < 0.01$, {***}\,$p < 0.001$, ns\,$p \geq 0.05$. Full protocol details are in Appendix~\ref{app:experimental_details}.

\subsection{Offline Decoder Comparison}
\label{subsec:offline}

\begin{figure*}[!t]
    \centering
    \subfloat[MC Maze]{%
        \includegraphics[width=0.42\textwidth]{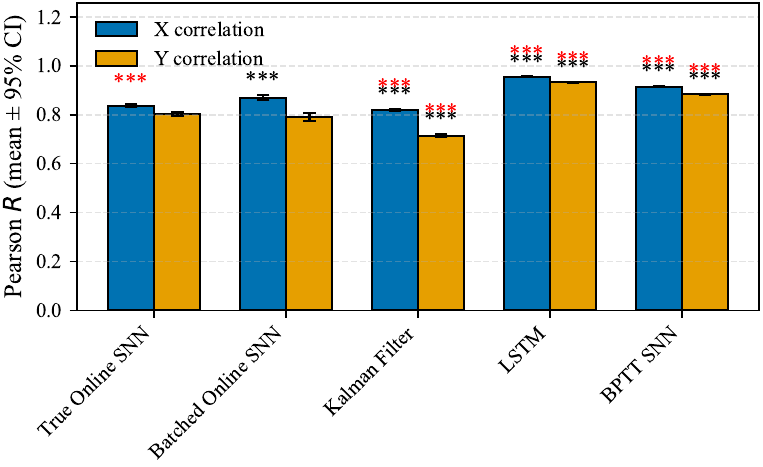}%
        \label{fig:mcmaze_comp}}
    \hfil
    \subfloat[Zenodo Indy]{%
        \includegraphics[width=0.42\textwidth]{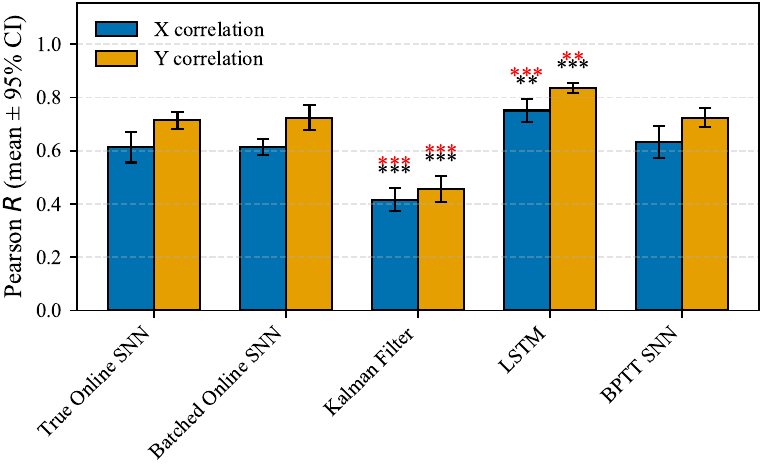}%
        \label{fig:zenodo_comp}}
    \caption{Decoder performance: Pearson $R$, mean $\pm$ SEM, $n{=}10$. Bar order: True Online SNN, Batched Online SNN, Kalman Filter, LSTM, BPTT SNN. Red: Welch's $t$-test vs.\ Batched Online SNN. Black: vs.\ True Online SNN ({*}\,$p{<}.05$, {**}\,$p{<}.01$, {***}\,$p{<}.001$, ns\,$p{\geq}.05$).}
    \label{fig:decoder_comparison}
\end{figure*}

Performance on MC~Maze and Zenodo~Indy is shown in Fig.~\ref{fig:decoder_comparison}. The Online SNN achieves Pearson correlations of $R \geq 0.81$ on MC~Maze and $R \geq 0.63$ on Zenodo~Indy. LSTM and GRU, trained with Adam and BPTT, attain higher offline correlations ($R \approx 0.79$ and $0.76$ on Zenodo); the gap reflects the absence of adaptive optimizer state and multi-step gradient flow in the online rule (Section~\ref{sec:discussion}). The BPTT-SNN, sharing the same architecture but trained with Adam, exceeds the Online SNN by ${\sim}8\%$ on MC~Maze ($R \approx 0.90$ vs.\ $0.82$) and ${\sim}2\%$ on Zenodo ($R \approx 0.68$ vs.\ $0.66$), confirming that temporal gradient flow provides an offline accuracy advantage on well-aligned datasets. Table~\ref{tab:baselines} summarizes decoder properties.

\begin{table}[!t]
\centering
\caption{Zenodo Decoder Comparison (Ten Sessions)}
\label{tab:baselines}
\resizebox{\columnwidth}{!}{%
\begin{tabular}{l c c c c}
\toprule
\textbf{Method} & \textbf{Avg corr} & \textbf{Online ($T{=}1$)} & \textbf{Optimizer} & \textbf{Hardware} \\
\midrule
Kalman & $0.44 \pm 0.06$ & Yes & None & CPU \\
MLP & $0.39 \pm 0.15$ & No & Adam & CPU/GPU \\
\textbf{SNN (ours)} & $\mathbf{0.66 \pm 0.05}$ & \textbf{Yes} & \textbf{Local rule} & \textbf{GPU} \\
GRU & $0.76 \pm 0.05$ & No & Adam + BPTT & GPU \\
LSTM & $0.79 \pm 0.04$ & No & Adam + BPTT & GPU \\
\bottomrule
\end{tabular}}
\end{table}

\begin{figure}[!t]
    \centering
    \subfloat[MC Maze]{%
        \includegraphics[width=0.48\columnwidth]{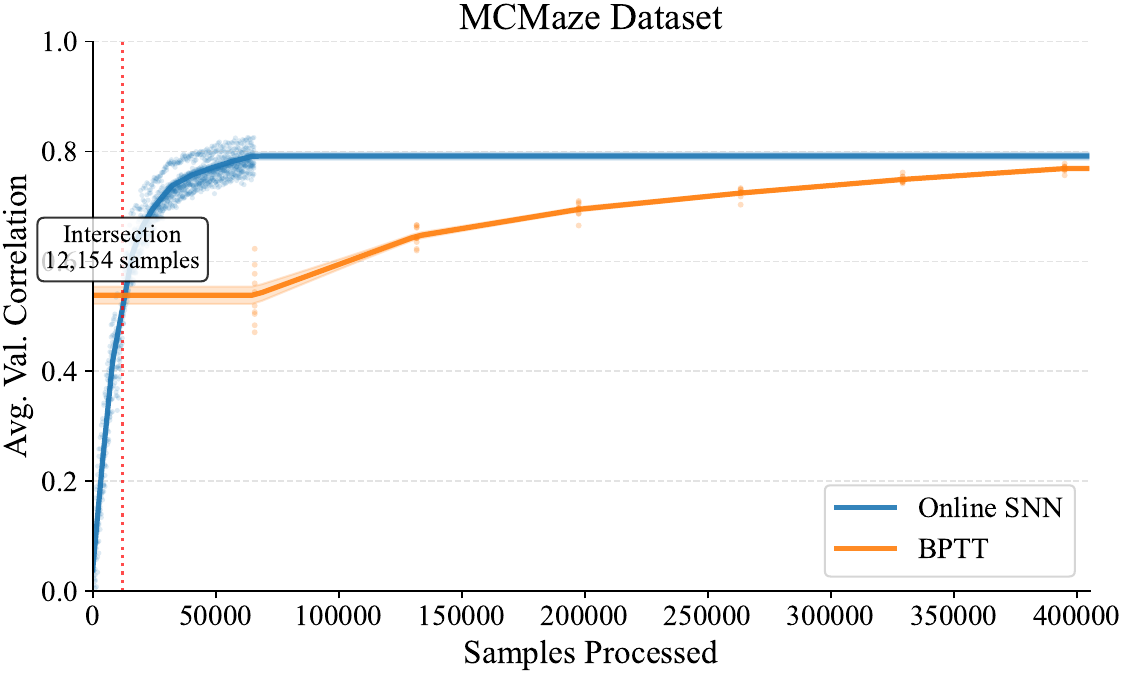}%
        \label{fig:mcmaze_lc}}
    \hfil
    \subfloat[Zenodo Indy]{%
        \includegraphics[width=0.48\columnwidth]{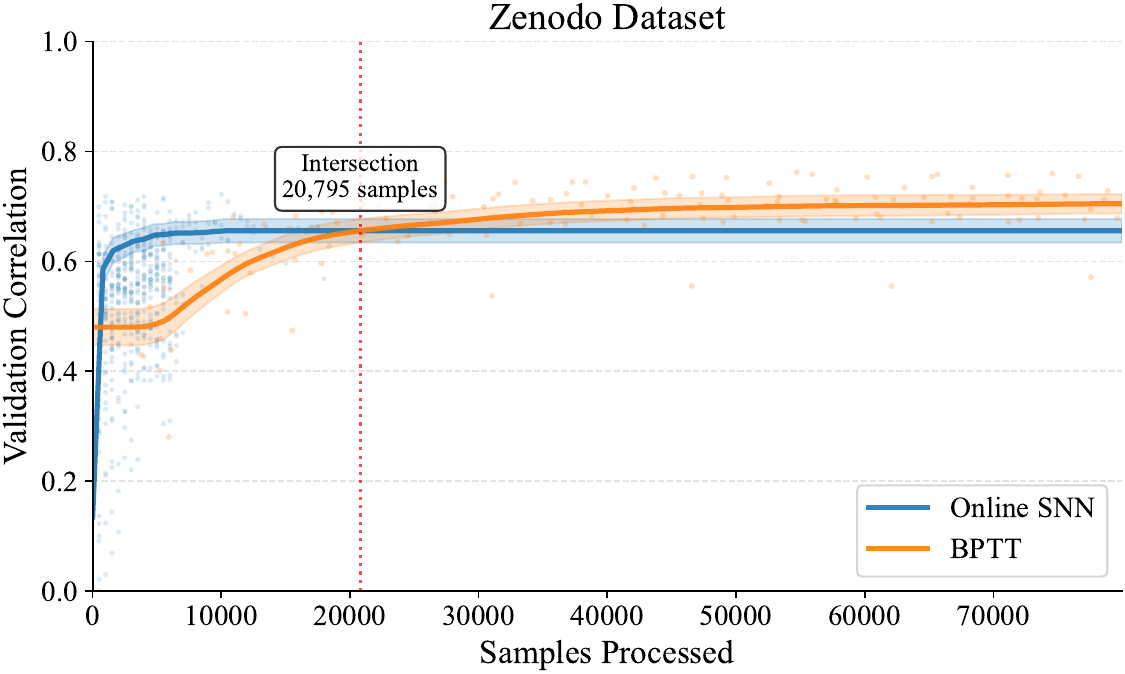}%
        \label{fig:zenodo_lc}}
    \caption{Learning curves: validation $R$ vs.\ samples processed. Mean $\pm$ SEM, $n{=}10$.}
    \label{fig:learning_curves}
\end{figure}

\textbf{Learning dynamics.}
Fig.~\ref{fig:learning_curves} compares learning curves on both datasets. The Online SNN reaches strong validation correlations after a few passes through the data despite an effective batch size of one. The BPTT-SNN attains a higher final $R$ when trained for many more epochs but requires substantially more optimizer steps and data exposure.

\subsection{Ablation Studies}
\label{subsec:ablations}

\begin{figure*}[!t]
    \centering
    \subfloat[Three-factor vs.\ $\Delta$-rule]{%
        \includegraphics[width=0.19\textwidth]{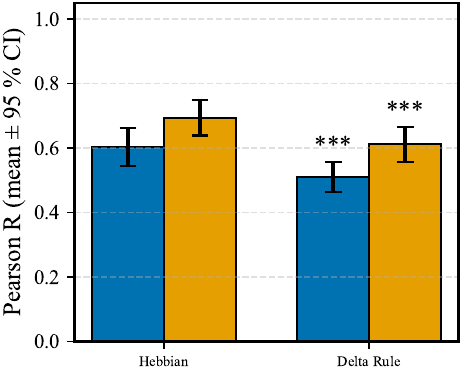}}
    \hfil
    \subfloat[Recurrent vs.\ feedforward]{%
        \includegraphics[width=0.19\textwidth]{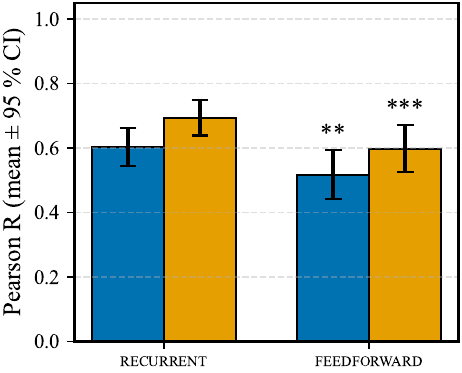}}
    \hfil
    \subfloat[Full vs.\ partial vs.\ no RMS]{%
        \includegraphics[width=0.19\textwidth]{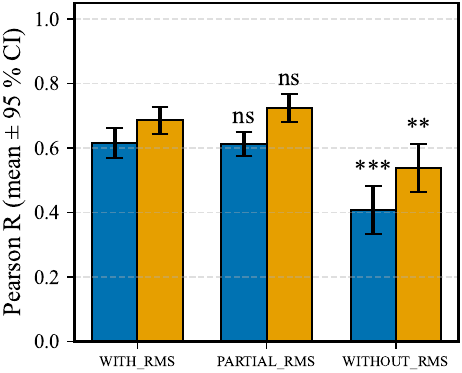}}
    \hfil
    \subfloat[Single vs.\ multi-timescale traces]{%
        \includegraphics[width=0.19\textwidth]{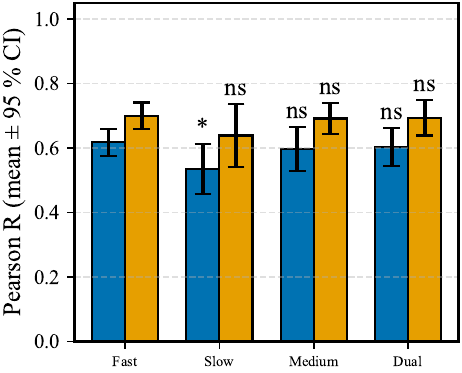}}
    \hfil
    \subfloat[Single vs.\ multi-timescale updates]{%
        \includegraphics[width=0.19\textwidth]{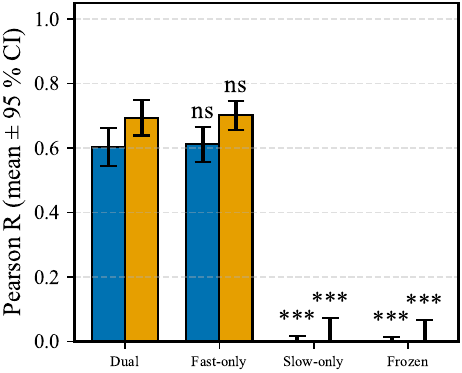}}
    \caption{Zenodo Indy ablation study. (d) Trace order: Fast, Slow, Medium, Dual. (e) Update order: Dual, Fast-only, Slow-only, Frozen. MC~Maze ablations are provided in Appendix~\ref{app:mcmaze_ablation}.}
    \label{fig:ablation_zenodo}
\end{figure*}

Component-wise ablations on Zenodo~Indy are shown in Fig.~\ref{fig:ablation_zenodo}; MC~Maze ablations appear in Appendix~\ref{app:mcmaze_ablation}. Table~\ref{tab:ablation_summary} quantifies the contribution of each component relative to a simplified delta-rule baseline.

\begin{table}[!t]
\centering
\caption{Component Contributions Relative to a Delta-Rule Baseline}
\label{tab:ablation_summary}
\resizebox{\columnwidth}{!}{%
\begin{tabular}{l l c c}
\toprule
\textbf{Component} & \textbf{Feature} & \textbf{Zenodo gain} & \textbf{MC~Maze gain} \\
\midrule
RMS normalization & Bitshift RMS EMAs & $+27.2$\% & $+0.8$\% \\
Recurrence & Recurrent layer & $+14.1$\% & $+0.7$\% \\
Hebbian accumulation & Three-factor traces & $+13.5$\% & $-0.1$\% \\
\bottomrule
\end{tabular}}
\end{table}

The dataset-dependent pattern reflects differences in temporal alignment and noise characteristics. MC~Maze applies an 80\,ms kinematic lag that improves signal-to-noise, reducing the need for fine-grained sensitivity gating: error signals already correlate well with presynaptic drive, so the delta rule approximates the three-factor rule. Zenodo~Indy provides continuous, self-paced reaches without explicit trial alignment or standardized lag, producing temporally broader error signals. In this setting, the surrogate gradient gate $d_{\text{LIF}}$ sharpens credit assignment by confining plasticity to moments when postsynaptic neurons are near threshold, and RMS homeostasis keeps update magnitudes stable across sessions with varying noise and sorting quality.

\textbf{Trace timescales.} A single fast or slow trace is optimal for one dataset but suboptimal for the other. The dual configuration yields near-optimal performance on both without dataset-specific tuning ($R{=}0.649$ on Zenodo, $R{=}0.767$ on MC~Maze).

\textbf{Synapse-specific vs.\ factorized traces.}
Table~\ref{tab:factorization} compares synapse-specific $O(N^2)$ traces against neuron-level $O(N)$ factorization on 10 Zenodo sessions. Synapse-specific traces yield approximately 0.06--0.08 higher correlation on both velocity axes under one-sided Welch $t$-tests (all $p < 0.04$). Intracortical electrodes sample neurons with heterogeneous temporal dynamics (fast-spiking interneurons vs.\ slow pyramidal cells, phasic velocity vs.\ tonic position encoding). Synapse-specific traces allow different downstream targets of the same electrode to maintain distinct fast/slow mixtures---a degree of temporal expressiveness that neuron-level factorization, which forces all outgoing connections from one input to share a single trace, cannot provide. The $O(N^2)$ cost remains practical at Utah-array scales (96--182 input channels).

\begin{table}[!t]
\centering
\caption{Synapse-Specific $O(N^2)$ vs.\ Factorized $O(N)$ Traces on Zenodo (10 Sessions)}
\label{tab:factorization}
\resizebox{\columnwidth}{!}{%
\begin{tabular}{l c c c c c}
\toprule
\textbf{Mode} & \textbf{Axis} & \textbf{$O(N^2)$} & \textbf{$O(N)$} & \textbf{$\Delta R$} & \textbf{$p$} \\
\midrule
True Online & X & $0.613 \pm 0.092$ & $0.537 \pm 0.084$ & $0.075$ & $0.036$ \\
True Online & Y & $0.714 \pm 0.053$ & $0.657 \pm 0.050$ & $0.057$ & $0.012$ \\
Batched Online & X & $0.614 \pm 0.048$ & $0.537 \pm 0.084$ & $0.076$ & $0.012$ \\
Batched Online & Y & $0.725 \pm 0.078$ & $0.657 \pm 0.050$ & $0.067$ & $0.018$ \\
\bottomrule
\end{tabular}}
\end{table}

\subsection{Closed-Loop Adaptation}
\label{subsec:closedloop}

\begin{figure*}[!t]
    \centering
    \subfloat[Remapping (90\%)]{%
        \includegraphics[width=0.24\textwidth]{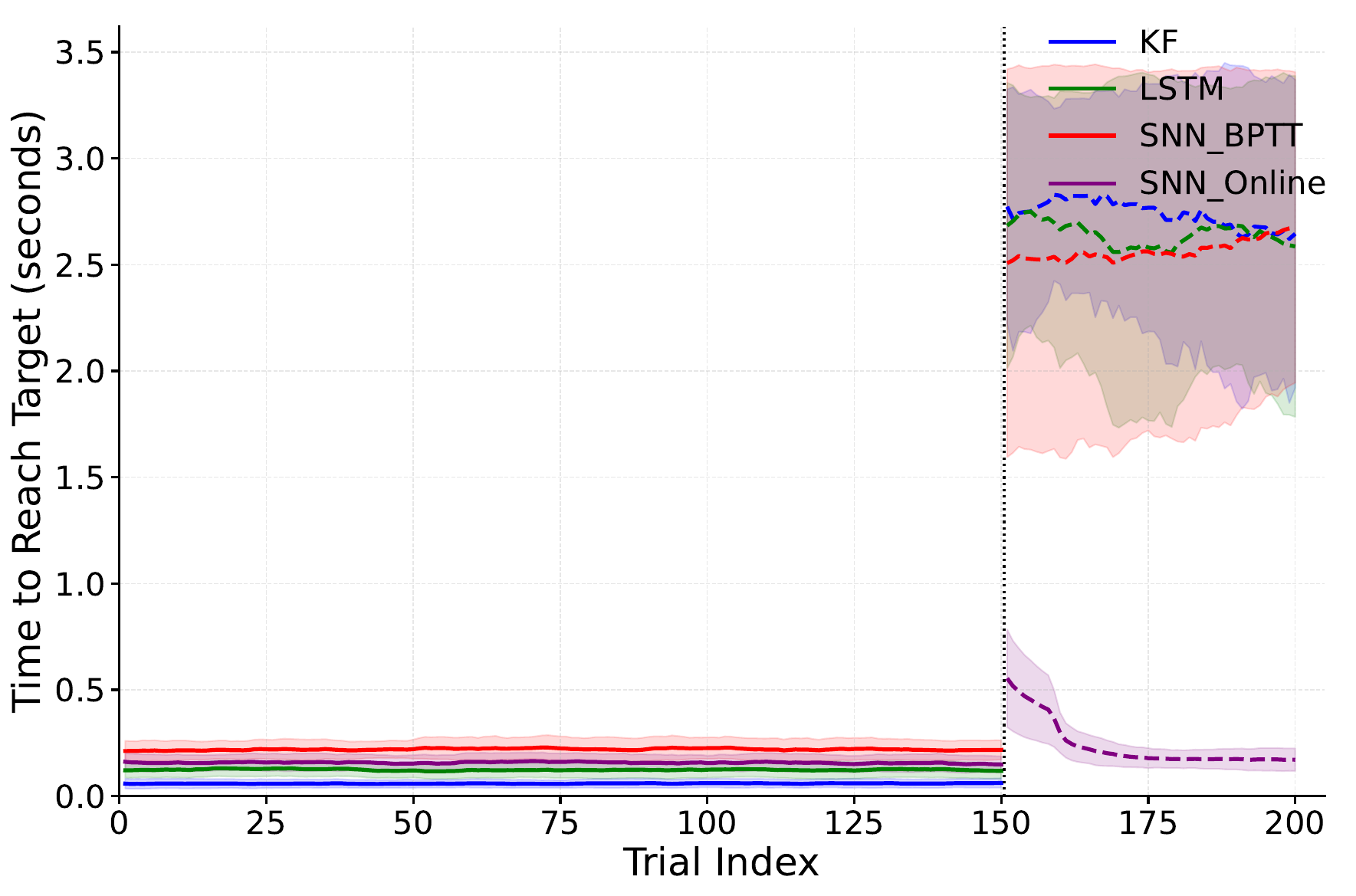}%
        \label{fig:remap}}
    \hfil
    \subfloat[Drift (90\%)]{%
        \includegraphics[width=0.24\textwidth]{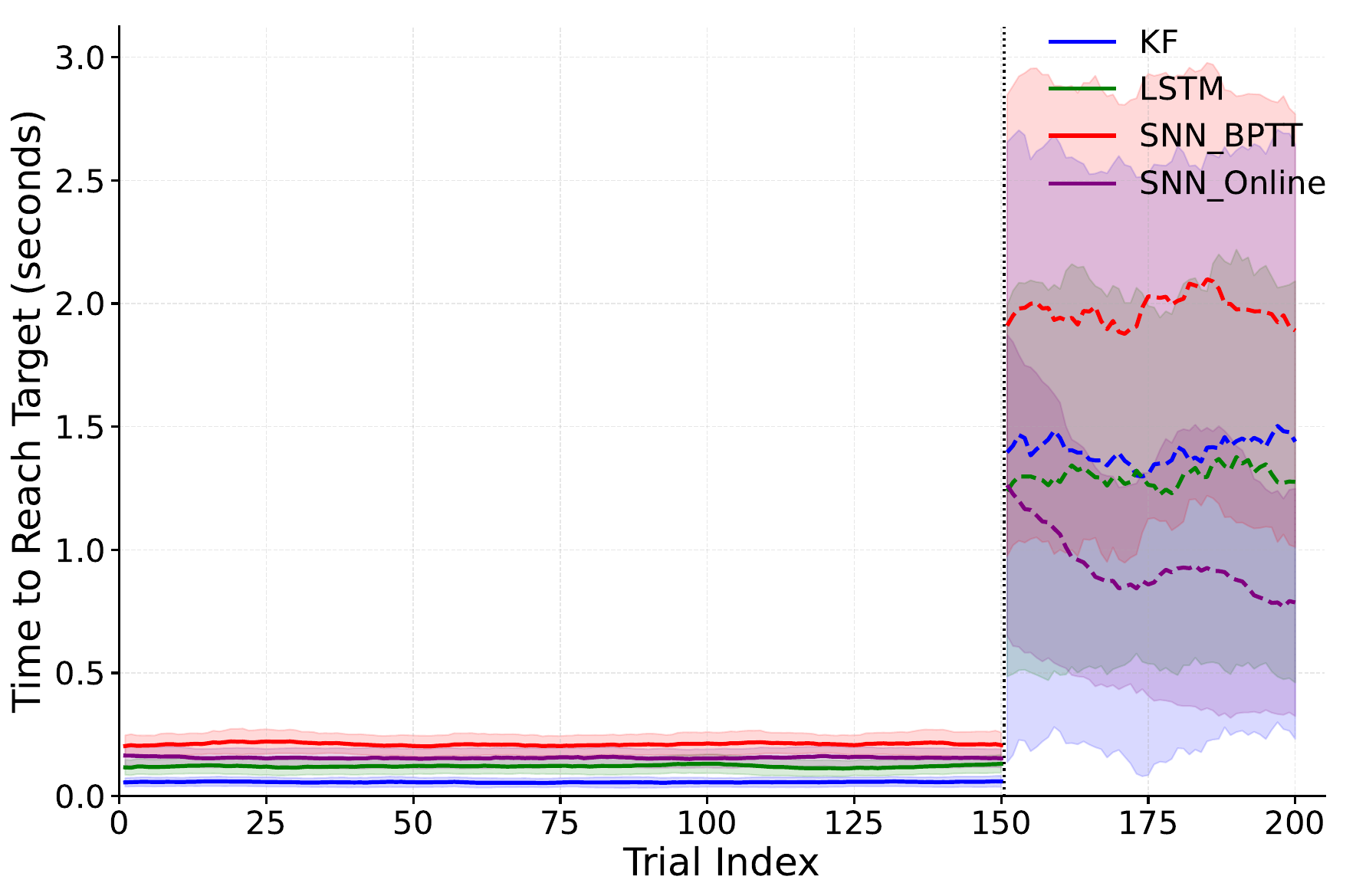}%
        \label{fig:drift}}
    \hfil
    \subfloat[Dropout (90\%)]{%
        \includegraphics[width=0.24\textwidth]{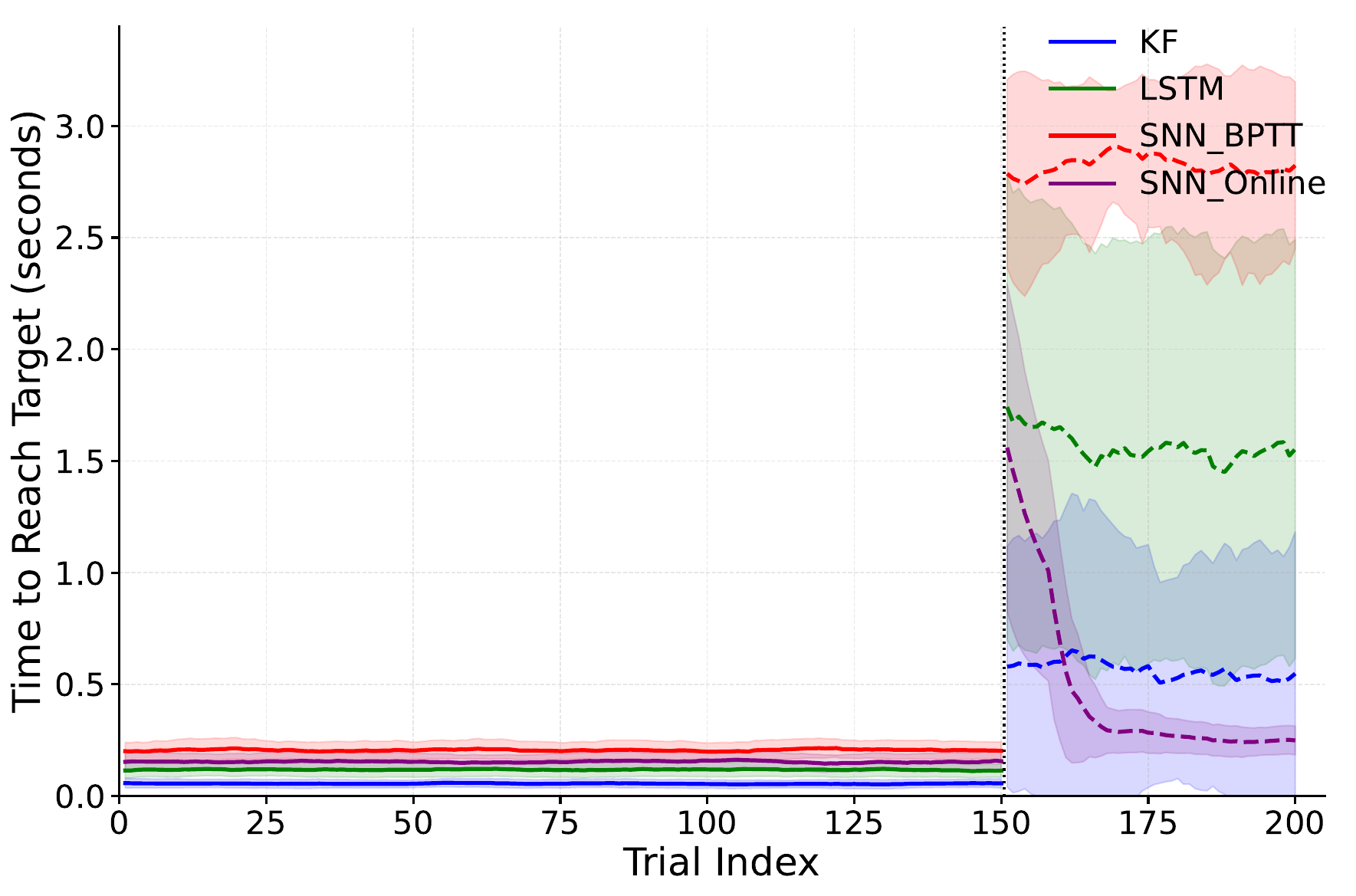}%
        \label{fig:dropout}}
    \hfil
    \subfloat[Online vs.\ offline calibration]{%
        \includegraphics[width=0.24\textwidth]{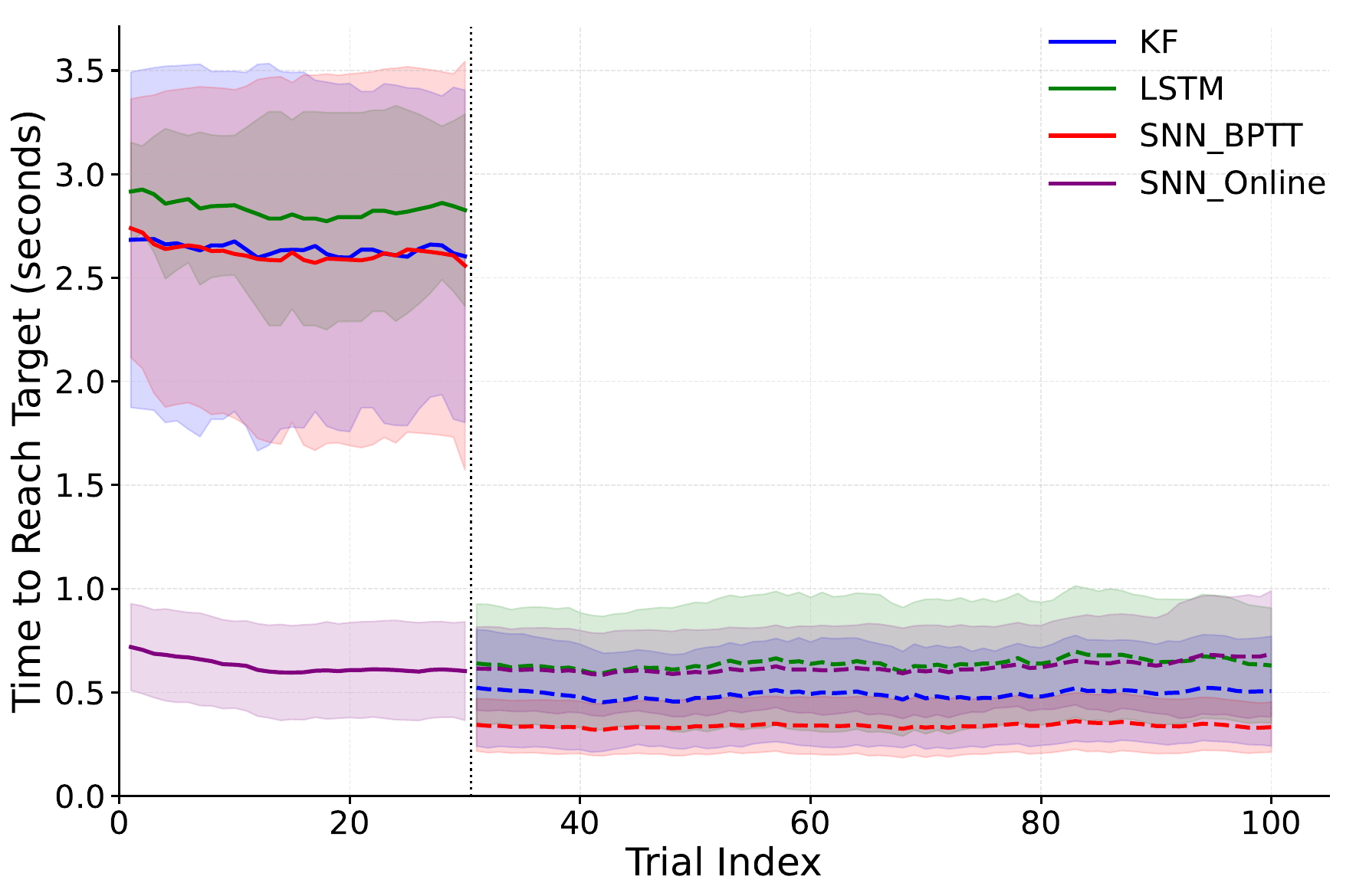}%
        \label{fig:nopretrain}}
    \caption{Closed-loop adaptation to disruptions (a--c) and learning from scratch (d). Mean over 10 runs. Full simulation methodology is provided in Appendix~\ref{app:closedloop_methods}.}
    \label{fig:closedloop}
\end{figure*}

Closed-loop simulations use synthetic 96-neuron populations with cosine-tuned directional preferences driving a 2D cursor control task~\cite{Cunningham2011}. Three disruptions mirror common intracortical failure modes: 90\% remapping (preferred direction reassignment), 90\% drift (firing rate compression with elevated baseline), and 90\% neuron dropout (Fig.~\ref{fig:closedloop}a--c).

Before disruption (first 150 reaches), all decoders---KF, LSTM, BPTT-SNN, and Online SNN---achieve mean time-to-target below 0.3\,s. After disruption onset, the Online SNN recovers to near pre-disruption performance within 15--20 reaches across all three conditions. Fixed-parameter decoders remain impaired: LSTM and BPTT-SNN reach times stay above 1.5\,s, while the Kalman filter degrades substantially under remapping and drift. The fast eligibility traces emphasize recent three-factor updates to correct for abrupt changes, slow traces preserve pre-disruption structure, and RMS homeostasis keeps update magnitudes well-scaled as firing statistics shift.

\textbf{Learning from scratch.}
In a separate protocol (Fig.~\ref{fig:closedloop}d), decoders are initialized with random weights and given access only to self-generated data. During the first 30 reaches, only the Online SNN updates per-timestep, improving from ${\sim}0.75$\,s to ${\sim}0.6$\,s, while fixed-weight decoders produce frequent timeouts at the 3\,s limit. After offline calibration on collected data, the other decoders reach similar time-to-target, but they require batch retraining with Adam on the accumulated data and provide no utility during the 30-reach collection phase itself.

\section{Discussion}
\label{sec:discussion}

\textbf{Offline--online trade-off.}
The 13--18\% offline accuracy gap relative to LSTM/GRU (Section~\ref{subsec:offline}) has two concrete sources. First, the Online SNN uses no per-parameter adaptive optimizer. Adam and RMSProp maintain $2P$ floating-point buffers---per-weight first- and second-moment estimates---that let each of the $P$ parameters receive an independently adapted step size. The Online SNN instead applies a single scalar learning rate per layer; the eligibility traces and momentum-smoothed slow accumulator (Section~\ref{subsec:learning_rule}) provide temporal smoothing of the update direction but do not adapt magnitudes per weight. Mathematically, each fast update is equivalent to a single-timestep gradient step on the instantaneous loss with a fixed learning rate, while the slow update applies global (not per-parameter) RMS normalization to the accumulated trace. This absence of per-weight adaptation is what separates the method from standard deep learning optimizers. Second, the Online SNN never unrolls a computational graph across time: it drops the temporal gradient entirely, losing the multi-step credit assignment that BPTT provides. This eliminates the $O(T)$ activation storage required by BPTT but sacrifices the ability to assign credit across timesteps. Both costs are unavoidable under the constraint of per-timestep, constant-memory operation on neuromorphic hardware within milliwatt power envelopes. In closed-loop deployment, the trade-off reverses: fixed-parameter decoders cannot adapt after neural disruptions, whereas the Online SNN recovers within 15--20 reaches (Section~\ref{subsec:closedloop}).

\textbf{Clinical relevance.}
Current intracortical BMI systems deployed in human trials (BrainGate, \cite{Gilja2015}) rely on Kalman-family decoders that require daily recalibration sessions, imposing a significant burden on users and clinicians~\cite{Brandman2018}. The Online SNN exceeds Kalman filter performance by a substantial margin ($+0.22$ mean correlation on Zenodo) while providing continuous per-timestep adaptation that does not interrupt ongoing use. As discussed above, the offline accuracy gap relative to LSTM/GRU becomes less relevant in deployment scenarios where fixed-parameter decoders degrade after neural disruptions and where the hardware required to run BPTT is unavailable. Translation to clinical use requires chronic nonhuman primate validation with longitudinal multi-session recordings, followed by human feasibility trials under an investigational device exemption.

\textbf{Scalability.}
The synapse-specific $O(N^2)$ traces that yield 0.06--0.08 higher correlation than $O(N)$ factorization (Table~\ref{tab:factorization}) are tractable at current Utah-array scales (96--182 channels) but may require factorized approximations as electrode counts increase in future high-density arrays.

\textbf{Comparison with recent SNN decoders.}
Table~\ref{tab:snn_comparison} positions this method relative to recent SNN-based intracortical decoders. It is the only method combining per-timestep online updates, dual-timescale traces, and an integer-friendly implementation.

\begin{table}[!t]
\centering
\caption{Comparison With Recent SNN Intracortical Decoders}
\label{tab:snn_comparison}
\resizebox{\columnwidth}{!}{%
\begin{tabular}{l c c c c c}
\toprule
\textbf{Method} & \textbf{Online} & \textbf{Dual-$\tau$} & \textbf{Int.} & \textbf{Perf.} & \textbf{Dataset} \\
\midrule
Taeckens~\cite{Taeckens2024} & Yes & No & No & $\rho{\approx}0.70$ & CRCNS, MC Maze \\
MotorSRNN~\cite{LiuMotorSRNN2023} & No & No & No & Classif. & M1 \\
LSS-CA~\cite{FuLSSCASNN2024} & No & No & No & $R^2{=}.50$--$.67$ & Rhesus \\
Adapt.\ Prun.~\cite{RivelliAdaptivelyPruned2024} & No & No & Part. & $R^2{=}.80$--$.98$ & NeuroBench \\
\textbf{Ours} & \textbf{Yes} & \textbf{Yes} & \textbf{Yes} & $\boldsymbol{\rho{=}.67}$--$\boldsymbol{.82}$ & Zenodo, MC~Maze \\
\bottomrule
\end{tabular}}
\end{table}

\textbf{Limitations.}
The learning rule is temporally local but spatially non-local: hidden-layer errors require weight-transpose propagation across layers (Section~\ref{subsec:learning_rule}), imposing symmetric forward and feedback paths~\cite{Lillicrap2020}. On-chip implementations exist~\cite{Renner2024}, but replacing $W^{\!\top}$ with spatially local feedback (e.g., feedback alignment) remains open. Closed-loop evaluation uses synthetic neural populations; validation on chronic recordings remains necessary. All training and memory measurements are obtained from PyTorch on GPU, not neuromorphic hardware; porting to fixed-point neuromorphic processors is future work. Hyperparameters ($K$, $\tau_{\text{fast}}$, $\tau_{\text{slow}}$, LIF decay) are tuned per dataset. Comparison is limited to standard BMI baselines; direct comparison with recent temporally local rules on shared intracortical benchmarks is left to future work.

\textbf{Future work.}
Validation on neuromorphic hardware (e.g., Loihi~\cite{Davies2018}), chronic longitudinal evaluation with multi-session recordings, and extension to higher-dimensional output spaces.

\section{Conclusion}
\label{sec:conclusion}

The experiments demonstrate that replacing BPTT with dual-timescale Hebbian accumulation incurs a moderate offline accuracy cost but unlocks per-timestep adaptation that no fixed-parameter decoder can match after neural disruptions. Ablations reveal that the relative importance of each component---three-factor gating, RMS homeostasis, and dual-timescale traces---depends on the noise and alignment characteristics of the recording, with the dual-timescale configuration providing robust cross-dataset performance without per-dataset tuning. Whether these properties hold on neuromorphic hardware and under chronic signal drift in human implants remains the central open question.

\section*{Data and Code Availability}
Both datasets are publicly available under CC-BY-4.0 licenses: MC~Maze through DANDI~\cite{Churchland2022} and Zenodo~Indy through Zenodo~\cite{ODoherty2017Zenodo}. Source code will be released upon publication.

\bibliographystyle{IEEEtran}
\bibliography{iclr2025_conference}

\begin{thebibliography}{10}
\providecommand{\url}[1]{#1}
\csname url@samestyle\endcsname
\providecommand{\newblock}{\relax}
\providecommand{\bibinfo}[2]{#2}
\providecommand{\BIBentrySTDinterwordspacing}{\spaceskip=0pt\relax}
\providecommand{\BIBentryALTinterwordstretchfactor}{4}
\providecommand{\BIBentryALTinterwordspacing}{\spaceskip=\fontdimen2\font plus
\BIBentryALTinterwordstretchfactor\fontdimen3\font minus \fontdimen4\font\relax}
\providecommand{\BIBforeignlanguage}[2]{{%
\expandafter\ifx\csname l@#1\endcsname\relax
\typeout{** WARNING: IEEEtran.bst: No hyphenation pattern has been}%
\typeout{** loaded for the language `#1'. Using the pattern for}%
\typeout{** the default language instead.}%
\else
\language=\csname l@#1\endcsname
\fi
#2}}
\providecommand{\BIBdecl}{\relax}
\BIBdecl

\bibitem{Pandarinath2017}
C.~Pandarinath \emph{et~al.}, ``High performance communication by people with paralysis using an intracortical brain--computer interface,'' \emph{eLife}, vol.~6, p. e18554, 2017.

\bibitem{Collinger2013}
J.~L. Collinger, B.~Wodlinger, J.~E. Downey, W.~Wang, E.~C. Tyler-Kabara, D.~J. Weber, A.~J.~C. McMorland, M.~Velliste, M.~L. Boninger, and A.~B. Schwartz, ``High-performance neuroprosthetic control by an individual with tetraplegia,'' \emph{The Lancet}, vol. 381, no. 9866, pp. 557--564, 2013.

\bibitem{Bouton2016}
C.~E. Bouton, A.~Shaikhouni, N.~V. Annetta, M.~A. Bockbrader, D.~A. Friedenberg, D.~M. Nielson, G.~Sharma, P.~B. Sederberg, T.~B. Glenn, W.~J. Mysiw, A.~G. Morgan, M.~Deogaonkar, and A.~R. Rezai, ``Restoring cortical control of functional movement in a human with quadriplegia,'' \emph{Nature}, vol. 533, no. 7602, pp. 247--250, 2016.

\bibitem{Dangi2014}
S.~Dangi, S.~Gowda, H.~G. Moorman, A.~L. Orsborn, K.~So, M.~Shanechi, and J.~M. Carmena, ``Continuous closed-loop decoder adaptation with a recursive maximum likelihood algorithm allows for rapid performance acquisition in brain--machine interfaces,'' \emph{Neural Computation}, vol.~26, no.~9, pp. 1811--1839, 2014.

\bibitem{Ganguly2009}
K.~Ganguly and J.~M. Carmena, ``Emergence of a stable cortical map for neuroprosthetic control,'' \emph{PLoS Biology}, vol.~7, no.~7, p. e1000153, 2009.

\bibitem{Sussillo2016}
D.~Sussillo, S.~D. Stavisky, J.~C. Kao, S.~I. Ryu, and K.~V. Shenoy, ``Making brain--machine interfaces robust to future neural variability,'' \emph{Nature Communications}, vol.~7, p. 13749, 2016.

\bibitem{Woeppel2021}
K.~A. Woeppel, J.~Li, B.~Tahayori, N.~A. Alba, and M.~I. Romero-Ortega, ``Recent advances in neural electrode--tissue interfaces,'' \emph{Current Opinion in Biomedical Engineering}, vol.~18, p. 100272, 2021.

\bibitem{Maynard1997}
E.~M. Maynard, C.~T. Nordhausen, and R.~A. Normann, ``The utah intracortical electrode array: A recording structure for potential brain--computer interfaces,'' \emph{Electroencephalography and Clinical Neurophysiology}, vol. 102, no.~3, pp. 228--239, 1997.

\bibitem{Huang2021}
X.~Huang, Y.~Xu, J.~Hua, W.~Yi, H.~Yin, R.~Hu, and S.~Wang, ``A review on signal processing approaches to reduce calibration time in eeg-based brain--computer interface,'' \emph{Frontiers in Neuroscience}, vol.~15, p. 733546, 2021.

\bibitem{Premchand2020}
B.~Premchand, K.~K. Toe, C.~Wang, S.~Shaikh, C.~Libedinsky, K.~K. Ang, and R.~Q. So, ``Decoding movement direction from cortical microelectrode recordings using an lstm-based neural network,'' in \emph{Proceedings of the 42nd Annual International Conference of the IEEE Engineering in Medicine \& Biology Society (EMBC)}.\hskip 1em plus 0.5em minus 0.4em\relax Montreal, QC, Canada: IEEE, 2020, pp. 3007--3010.

\bibitem{Wu2004}
W.~Wu, M.~J. Black, D.~Mumford, Y.~Gao, E.~Bienenstock, and J.~P. Donoghue, ``Modeling and decoding motor cortical activity using a switching kalman filter,'' \emph{IEEE Transactions on Biomedical Engineering}, vol.~51, no.~6, pp. 933--942, 2004.

\bibitem{Jiang2018}
J.~Jiang, R.~Xu, D.~Ming, B.~Wan, R.~Zhang, B.~Wang, and B.~Hu, ``A novel online multi-class adaptive brain--computer interface based on transfer learning,'' \emph{Journal of Neural Engineering}, vol.~15, no.~3, p. 036008, 2018.

\bibitem{Carmena2013}
J.~M. Carmena, ``Advances in neuroprosthetic learning and control,'' \emph{PLoS Biology}, vol.~11, no.~5, p. e1001561, 2013.

\bibitem{Orsborn2014}
A.~L. Orsborn, H.~G. Moorman, S.~A. Overduin \emph{et~al.}, ``Closed-loop decoder adaptation shapes neural plasticity for skillful neuroprosthetic control,'' \emph{Neuron}, vol.~82, no.~6, pp. 1380--1393, 2014.

\bibitem{Bhaduri2018}
A.~Bhaduri, A.~Banerjee, S.~Roy, S.~Kar, and A.~Basu, ``Spiking neural classifier with lumped dendritic nonlinearity and binary synapses: A current mode vlsi implementation and analysis,'' \emph{Neural Computation}, vol.~30, no.~3, pp. 723--760, 2018.

\bibitem{Basu2022SNNICReview}
A.~Basu, C.~Frenkel, L.~Deng, and X.~Zhang, ``Spiking neural network integrated circuits: A review of trends and future directions,'' \emph{arXiv preprint}, 2022, authors listed alphabetically.

\bibitem{Florian2007}
R.~V. Florian, ``Reinforcement learning through modulation of spike-timing-dependent synaptic plasticity,'' \emph{Neural Computation}, vol.~19, no.~6, pp. 1468--1502, 2007.

\bibitem{Fremaux2016}
N.~Fr{\'e}maux and W.~Gerstner, ``Neuromodulated spike-timing-dependent plasticity, and theory of three-factor learning rules,'' \emph{Frontiers in Neural Circuits}, vol.~9, p.~85, 2016.

\bibitem{Izhikevich2007}
E.~M. Izhikevich, ``Solving the distal reward problem through linkage of {STDP} and dopamine signaling,'' \emph{Cerebral Cortex}, vol.~17, no.~10, pp. 2443--2452, 2007.

\bibitem{Zenke2018}
F.~Zenke and S.~Ganguli, ``Superspike: Supervised learning in multilayer spiking neural networks,'' \emph{Neural Computation}, vol.~30, no.~6, pp. 1514--1541, 2018.

\bibitem{Bellec2020}
G.~Bellec, F.~Scherr, D.~Salaj \emph{et~al.}, ``A solution to the learning dilemma for recurrent networks of spiking neurons,'' \emph{Nature Communications}, vol.~11, p. 3625, 2020.

\bibitem{Xiao2022OTTT}
M.~Xiao, Q.~Meng, Z.~Zhang, D.~He, and Z.~Lin, ``Online training through time for spiking neural networks,'' in \emph{Advances in Neural Information Processing Systems}, vol.~35, 2022, pp. 20\,717--20\,730.

\bibitem{ApolinarioRoy2025STLLR}
M.~P.~E. Apolinario and K.~Roy, ``S-tllr: Stdp-inspired temporal local learning rule for spiking neural networks,'' \emph{Transactions on Machine Learning Research}, 2025.

\bibitem{Meng2023SLTT}
Q.~Meng, M.~Xiao, S.~Yan, Y.~Wang, Z.~Lin, and Z.-Q. Luo, ``Towards memory-and time-efficient backpropagation for training spiking neural networks,'' in \emph{Proceedings of the IEEE/CVF International Conference on Computer Vision}, 2023, pp. 6166--6176.

\bibitem{ApolinarioRoyFrenkel2025TESS}
M.~P.~E. Apolinario, K.~Roy, and C.~Frenkel, ``Tess: A scalable temporally and spatially local learning rule for spiking neural networks,'' \emph{arXiv preprint}, 2025.

\bibitem{Neftci2019}
E.~O. Neftci, H.~Mostafa, and F.~Zenke, ``Surrogate gradient learning in spiking neural networks: Bringing the power of gradient-based optimization to spiking neural networks,'' \emph{IEEE Signal Processing Magazine}, vol.~36, no.~6, pp. 51--63, 2019.

\bibitem{Renner2024}
A.~Renner, M.~Evanusa, R.~Gra{\c{c}}a, Y.~L{\"o}wenstein, S.~Vassanelli, D.~Bhatt, Y.~Lin, G.~Bhatt, O.~Rhodes, and E.~Neftci, ``The backpropagation algorithm implemented on spiking neuromorphic hardware,'' \emph{Nature Communications}, vol.~15, p. 7724, 2024.

\bibitem{Pfister2006}
J.-P. Pfister and W.~Gerstner, ``Triplets of spikes in a model of spike timing-dependent plasticity,'' \emph{Journal of Neuroscience}, vol.~26, no.~38, pp. 9673--9682, 2006.

\bibitem{Benna2016}
M.~K. Benna and S.~Fusi, ``Computational principles of synaptic memory consolidation,'' \emph{Nature Neuroscience}, vol.~19, no.~12, pp. 1697--1706, 2016.

\bibitem{Davies2018}
M.~Davies, N.~Srinivasa, T.-H. Lin \emph{et~al.}, ``Loihi: A neuromorphic manycore processor with on-chip learning,'' \emph{IEEE Micro}, vol.~38, no.~1, pp. 82--99, 2018.

\bibitem{Hassan2024}
A.~Hasssan, J.~Meng, A.~Anupreetham, and J.-s. Seo, ``{SpQuant-SNN}: Ultra-low precision membrane potential with sparse activations unlock the potential of on-device spiking neural networks applications,'' \emph{Frontiers in Neuroscience}, vol.~18, p. 1440000, 2024.

\bibitem{Churchland2022}
\BIBentryALTinterwordspacing
M.~Churchland and M.~Kaufman, ``Mc\_maze: macaque primary motor and dorsal premotor cortex spiking activity during delayed reaching (version 0.220113.0400) [data set],'' 2022. [Online]. Available: \url{https://doi.org/10.48324/dandi.000128/0.220113.0400}
\BIBentrySTDinterwordspacing

\bibitem{ODoherty2017Zenodo}
\BIBentryALTinterwordspacing
J.~E. O'Doherty, M.~M.~B. Cardoso, J.~G. Makin, and P.~N. Sabes, ``Nonhuman primate reaching with multichannel sensorimotor cortex electrophysiology,'' Dataset, 2017. [Online]. Available: \url{https://doi.org/10.5281/zenodo.583331}
\BIBentrySTDinterwordspacing

\bibitem{Cunningham2011}
J.~P. Cunningham, P.~Nuyujukian, V.~Gilja, C.~A. Chestek, S.~I. Ryu, and K.~V. Shenoy, ``A closed-loop human simulator for investigating the role of feedback control in brain--machine interfaces,'' \emph{Journal of Neurophysiology}, vol. 105, no.~4, pp. 1932--1949, 2011.

\bibitem{Gilja2015}
V.~Gilja, C.~Pandarinath, C.~H. Blabe, P.~Nuyujukian, J.~D. Simeral, A.~A. Sarma, B.~L. Sorice, J.~A. Perge, B.~Jarosiewicz, L.~R. Hochberg, K.~V. Shenoy, and J.~M. Henderson, ``Clinical translation of a high-performance neural prosthesis,'' \emph{Nature Medicine}, vol.~21, no.~10, pp. 1142--1145, 2015.

\bibitem{Brandman2018}
D.~M. Brandman, T.~Hosman, J.~Saab, M.~C. Burkhart, B.~E. Shanahan, J.~G. Ciancibello, A.~A. Sarma, D.~J. Milstein, C.~E. Vargas-Irwin, B.~Franco, J.~Kelemen, C.~Blabe, B.~A. Murphy, D.~R. Young, F.~R. Willett, C.~Pandarinath, S.~D. Stavisky, R.~F. Kirsch, B.~L. Walter, A.~B. Ajiboye, S.~S. Cash, E.~N. Eskandar, J.~P. Miller, J.~A. Sweet, K.~V. Shenoy, J.~M. Henderson, B.~Jarosiewicz, M.~T. Harrison, J.~D. Simeral, and L.~R. Hochberg, ``Rapid calibration of an intracortical brain--computer interface for people with tetraplegia,'' \emph{Journal of Neural Engineering}, vol.~15, no.~2, p. 026007, 2018.

\bibitem{Taeckens2024}
E.~A. Taeckens and S.~Shah, ``A spiking neural network with continuous local learning for robust online brain--machine interface,'' \emph{Journal of Neural Engineering}, vol.~20, no.~6, p. 066042, 2024.

\bibitem{LiuMotorSRNN2023}
T.~Liu, Y.~Chua, Y.~Ning, P.~Liu, Y.~Zhang, T.~Li, G.~Wan, Z.~Wan, W.~Chen, and S.~Zhang, ``{motorSRNN}: A spiking recurrent neural network inspired by brain topology for the effective and efficient decoding of cortical spike trains,'' \emph{Neural Networks}, 2023.

\bibitem{FuLSSCASNN2024}
H.~Fu, P.~Zhang, S.~Yang, H.~Zhang, Z.~Wang, and D.~Wu, ``Effective and efficient intracortical brain signal decoding with spiking neural networks,'' \emph{arXiv preprint}, 2024.

\bibitem{RivelliAdaptivelyPruned2024}
F.~Rivelli, M.~Popov, C.~S. Kouzinopoulos, and G.~Tang, ``Adaptively pruned spiking neural networks for energy-efficient intracortical neural decoding,'' \emph{arXiv preprint}, 2024.

\bibitem{Lillicrap2020}
T.~P. Lillicrap, A.~Santoro, L.~Marris, C.~J. Akerman, and G.~Hinton, ``Backpropagation and the brain,'' \emph{Nature Reviews Neuroscience}, vol.~21, no.~6, pp. 335--346, 2020.

\bibitem{Roy2019}
K.~Roy, A.~Jaiswal, and P.~Panda, ``Towards spike-based machine intelligence with neuromorphic computing,'' \emph{Nature}, vol. 575, no. 7784, pp. 607--617, 2019.

\bibitem{Furber2014}
S.~B. Furber, F.~Galluppi, S.~Temple, and L.~A. Plana, ``The spinnaker project,'' \emph{Proceedings of the IEEE}, vol. 102, no.~5, pp. 652--665, 2014.

\end{thebibliography}

\appendices

\section{Notation}
\label{app:notation}

\begin{table}[!t]
\centering
\caption{Key Notation and Symbols}
\label{tab:notation}
\begin{tabular}{cl}
\toprule
\textbf{Symbol} & \textbf{Description} \\
\midrule
$\mathbf{x}_t$ & Input spike count vector at time $t$ \\
$\mathbf{y}_t$ & Target 2D velocity at time $t$ \\
$\hat{\mathbf{y}}_t$ & Predicted 2D velocity at time $t$ \\
$\mathbf{u}^{(k)}_t$ & Membrane potentials for layer $k$ at time $t$ \\
$\mathbf{s}^{(k)}_t$ & Spike outputs for layer $k$ at time $t$ \\
$d_{\text{LIF}}$ & Fast sigmoid surrogate gradient, Eq.~(S) \\
$E^{\text{fast}}, E^{\text{slow}}$ & Fast and slow eligibility traces \\
$\lambda_{\text{fast}}, \lambda_{\text{slow}}$ & Decay coefficients: $e^{-\Delta t / \tau_{\text{fast/slow}}}$ \\
$\tau_{\text{fast}}, \tau_{\text{slow}}$ & Trace time constants (120\,ms, 700\,ms) \\
$\alpha_{\text{mix}}$ & Mixing coefficient for combining traces \\
$\eta_{\text{fast}}, \eta_{\text{slow}}$ & Fast and slow learning rates \\
$K$ & Consolidation window (timesteps) \\
$\mathcal{R}(\cdot)$ & RMS normalization function \\
\bottomrule
\end{tabular}
\end{table}

\section{Overview of Spiking Neural Networks}
\label{app:overviewSNN}

Spiking Neural Networks (SNNs) employ discrete, event-driven spikes for information transmission and processing, in contrast to traditional artificial neurons that process continuous-valued activations. This event-driven representation supports sparse computation and is well suited for neuromorphic hardware designed to leverage spike-based communication~\cite{Roy2019,Furber2014}. The Leaky Integrate-and-Fire (LIF) neuron model, used in this work, describes the membrane potential $u(t)$ as:
\[
\tau_m\,\frac{du(t)}{dt}
= -\bigl(u(t)-u_{\mathrm{rest}}\bigr) + R\,I(t),
\]
where $\tau_m$ is the membrane time constant, $u_{\mathrm{rest}}$ the resting potential, $R$ the input resistance, and $I(t)$ the synaptic input current. When $u(t)$ exceeds a threshold $u_{\mathrm{th}}$, the neuron emits a spike and $u(t)$ resets to $u_{\mathrm{reset}}$. This threshold-based dynamics can reduce computational load when spikes are sparse, making SNNs relevant for power-constrained deployment settings such as implantable BMIs.

SNNs are typically trained using learning rules that incorporate modulatory signals beyond simple pre/post-synaptic correlations. Three-factor rules~\cite{Fremaux2016,Zenke2018,Bellec2020} combine local Hebbian terms with a global error or reward signal, enabling supervised online learning without the need to unroll the network in time. The challenge of applying spike-based learning rules on conventional hardware can involve substantial overhead for tracking spike dynamics, motivating the integer-friendly implementation presented in Section~\ref{subsec:stability}.

\section{Experimental Details}
\label{app:experimental_details}

\subsection{Datasets and Preprocessing}

\textbf{MC~Maze.} Raw neural and kinematic data are resampled to 10\,ms resolution. For each trial, data are aligned to movement onset; kinematics are lagged by 80\,ms relative to spikes. Spikes and velocity are binned into 100\,ms windows with 10\,ms stride. Target velocities $(v_x, v_y)$ are z-score normalized using training set statistics. SNN inputs remain as raw spike counts; baselines receive rate-normalized inputs. Data splitting is performed on entire trials (70/15/15\%) to prevent temporal leakage.

\textbf{Zenodo~Indy.} Spike events are binned into 50\,ms non-overlapping windows as raw counts. Cursor velocity is smoothed using a 4th-order 10\,Hz Butterworth filter, differentiated, and averaged per bin. Zero kinematic lag is used. Target velocities are z-score normalized. Chronological splits (70/15/15\%) reflect real-world deployment where the decoder is trained on past data and tested on future data.

\subsection{Decoder Architectures}

In Batched Online mode, sequences are grouped into mini-batches of 32 for parallelism. In True Online mode, hidden states persist across the full chronological stream without reset between sequences. All neural network decoders are implemented in PyTorch and snnTorch, evaluated on a single A100 or H100 GPU over 10 independent runs per configuration.

\section{MC~Maze Ablation Results}
\label{app:mcmaze_ablation}

\begin{figure}[!t]
    \centering
    \subfloat[Three-factor vs.\ $\Delta$-rule]{%
        \includegraphics[width=0.31\columnwidth]{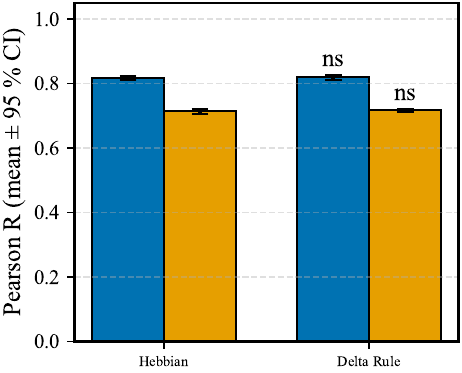}}
    \hfil
    \subfloat[Feedforward vs.\ recurrent]{%
        \includegraphics[width=0.31\columnwidth]{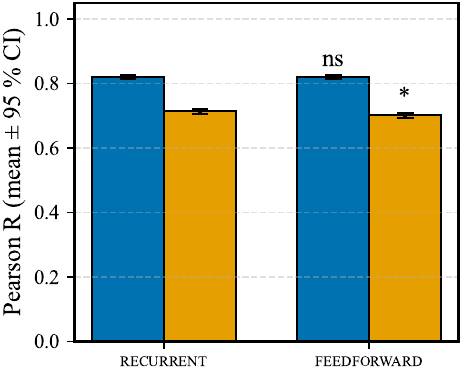}}
    \hfil
    \subfloat[RMS ablation]{%
        \includegraphics[width=0.31\columnwidth]{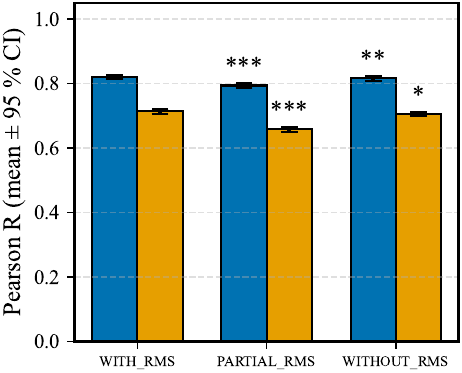}}
    \\[4pt]
    \subfloat[Multi-timescale traces]{%
        \includegraphics[width=0.31\columnwidth]{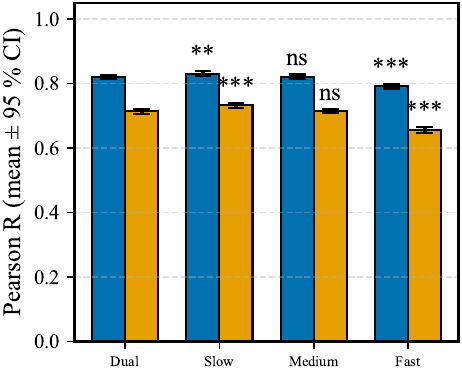}}
    \hfil
    \subfloat[Multi-timescale updates]{%
        \includegraphics[width=0.31\columnwidth]{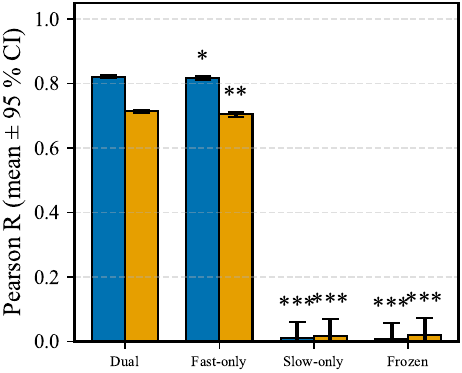}}
    \caption{MC~Maze ablation study. (d) Trace order: Fast, Slow, Medium, Dual. (e) Update order: Dual, Fast-only, Slow-only, Frozen.}
    \label{fig:ablation_mcmaze}
\end{figure}

On MC~Maze (Fig.~\ref{fig:ablation_mcmaze}), three-factor and delta rules perform nearly identically, as expected given the dataset-dependent pattern described in Section~\ref{subsec:ablations}. Recurrence provides a consistent but modest gain. Full RMS normalization is marginally beneficial; partial normalization degrades performance. Slow traces are strongest on this well-aligned dataset. Both fast and slow update pathways are necessary: slow-only or frozen updates collapse performance toward zero. Table~\ref{tab:timescale_ablation} shows that the dual-timescale configuration achieves near-optimal correlation on both datasets simultaneously.

\begin{table}[!t]
\centering
\caption{Dual-Timescale Eligibility Trace Robustness}
\label{tab:timescale_ablation}
\resizebox{\columnwidth}{!}{%
\begin{tabular}{l c c l}
\toprule
\textbf{Configuration} & \textbf{Zenodo} & \textbf{MC~Maze} & \textbf{Effect} \\
\midrule
$\tau_{\text{fast}}$ only (120\,ms) & $0.659$ & $0.724$ & Optimal on Zenodo \\
$\tau_{\text{slow}}$ only (700\,ms) & $0.588$ & $0.782$ & Optimal on MC~Maze \\
\textbf{Dual} & \textbf{0.649} & \textbf{0.767} & Near-optimal on both \\
\bottomrule
\end{tabular}}
\end{table}

\section{Closed-Loop Simulation Methods}
\label{app:closedloop_methods}

The closed-loop simulations model a 2D cursor control task on an $800 \times 600$ unit screen using a synthetic neural population of 96 neurons with cosine-tuned directional preferences uniformly distributed across four quadrants.

\textbf{Spike generation.} Given target velocity $\mathbf{v}(t)$, firing rate for neuron $i$ is:
\begin{equation}
r_i(t) = r_{\min} + (r_{\max} - r_{\min}) \cdot \max\!\left(0, \frac{\mathbf{d}_i \cdot \hat{\mathbf{v}}(t) + 0.5}{1.5}\right) \tag{S1}
\end{equation}
where $\hat{\mathbf{v}}(t) = \mathbf{v}(t) / (2\|\mathbf{v}(t)\|)$, $r_{\min}{=}5$\,Hz, $r_{\max}{=}100$\,Hz. Spike probability is $p_i(t) = \mathrm{clamp}(r_i(t) \cdot \Delta t + \mathcal{N}(0, \sigma^2), 0, 1)$ with $\Delta t{=}0.01$\,s, $\sigma{=}0.02$. Spikes are sampled as $s_i(t) \sim \mathrm{Bernoulli}(p_i(t))$.

\textbf{Control dynamics.} At each timestep, desired velocity is computed using a proportional controller toward the target. The decoder produces $\hat{\mathbf{v}}_{\text{decoder}}(t)$, which updates cursor position with movement scaling factor $\alpha{=}5$. Success is $\|\mathbf{p}_{\text{target}} - \mathbf{p}_{\text{cursor}}\| < R$ within 300 timesteps (3\,s).

\textbf{Pre-disruption calibration.} All decoders are first calibrated using 10,000 timesteps of closed-loop data generated by an untrained LSTM. The KF is fitted via maximum likelihood; the LSTM and BPTT-SNN are trained offline on this data (30 epochs, Adam, $\mathrm{lr}{=}10^{-3}$). The Online SNN is pre-trained through 100 reach attempts. After calibration, all decoders achieve mean time-to-target below 0.3\,s.

\textbf{Disruptions.} Remapping: a fraction of neurons receive reassigned preferred directions. Drift: $r_{\max}$ is compressed and $r_{\min}$ elevated. Dropout: a binary mask silences a fraction of neurons.

\textbf{No-pretrain protocol.} All decoders are initialized with random weights and each decoder independently controls the cursor for 30 reaches, generating its own closed-loop data. During these 30 reaches, only the Online SNN updates its weights per-timestep. Offline-trainable decoders (KF, LSTM, BPTT-SNN) are then calibrated on their respective collected data and all decoders are evaluated on 70 subsequent reaches.

Results are averaged over 10 independent runs with different random seeds.

\section{Memory Scaling Analysis}
\label{app:memory_scaling}

\begin{figure}[!t]
    \centering
    \subfloat[Duration (Zenodo)]{%
        \includegraphics[width=0.31\columnwidth]{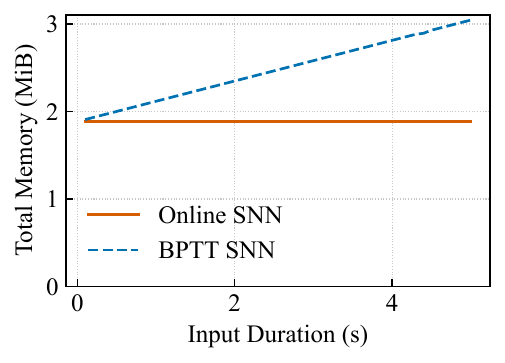}}
    \hfil
    \subfloat[Neurons (Zenodo)]{%
        \includegraphics[width=0.31\columnwidth]{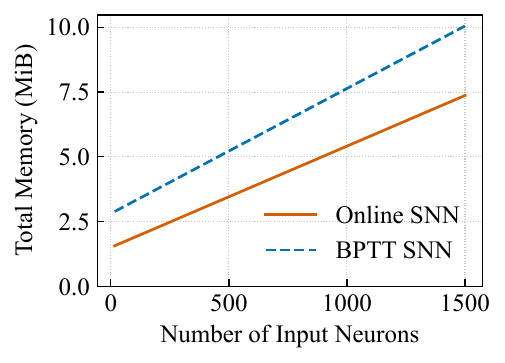}}
    \hfil
    \subfloat[Region map (Zenodo)]{%
        \includegraphics[width=0.31\columnwidth]{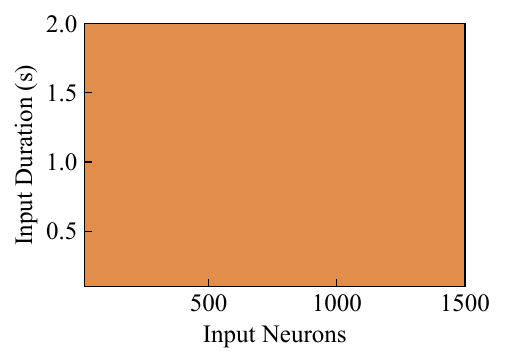}}
    \\[4pt]
    \subfloat[Duration (MC~Maze)]{%
        \includegraphics[width=0.31\columnwidth]{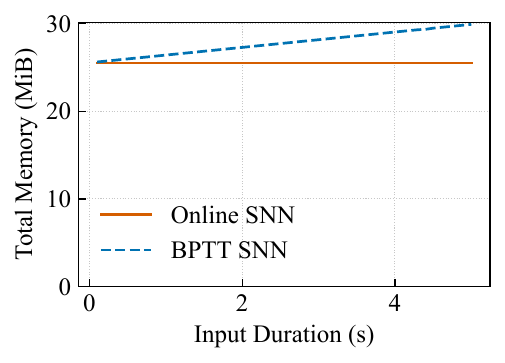}}
    \hfil
    \subfloat[Neurons (MC~Maze)]{%
        \includegraphics[width=0.31\columnwidth]{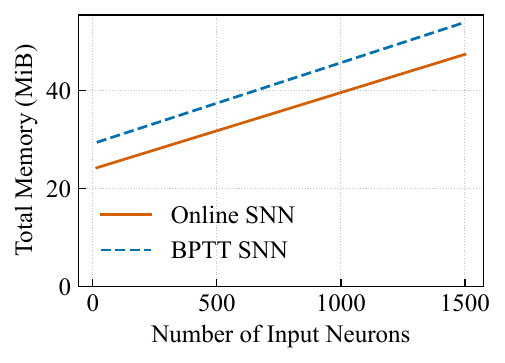}}
    \hfil
    \subfloat[Region map (MC~Maze)]{%
        \includegraphics[width=0.31\columnwidth]{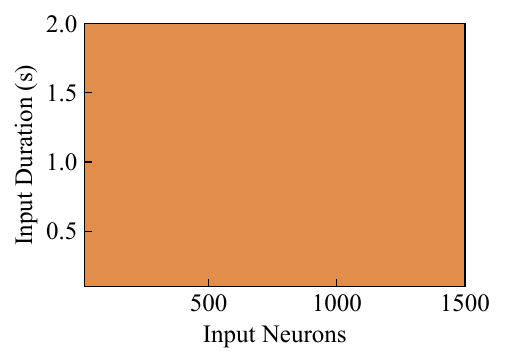}}
    \caption{Memory scaling. (a,d) Memory vs.\ duration: Online SNN constant-in-$T$. (b,e) Scaling with dimensionality. (c,f) Regions where Online SNN (orange) or BPTT (blue) has lower memory.}
    \label{fig:memory_scaling}
\end{figure}

Table~\ref{tab:memory_seq} shows that Online SNN peak memory remains constant while BPTT memory grows roughly linearly with $T$, leading to reductions of 24--97\% as sequence length increases from 50 to 5000. Table~\ref{tab:memory_hidden} shows that varying hidden width from 32 to 4096 neurons yields absolute savings from 2\,MB to 173\,MB. Fig.~\ref{fig:memory_scaling} visualizes these scaling properties for two representative architectures.

\begin{table}[!t]
\centering
\caption{Peak Memory (MB) vs.\ Sequence Length $T$}
\label{tab:memory_seq}
\resizebox{\columnwidth}{!}{%
\begin{tabular}{c c c c c c c}
\toprule
\textbf{Dataset} & \textbf{Hidden} & \textbf{SeqLen} & \textbf{Online} & \textbf{BPTT} & \textbf{Reduction} & \textbf{Savings} \\
\midrule
Zenodo & 256 & 50   & 1.88 & 2.46  & 23.7\% & 0.58 \\
MC~Maze & 1024 & 50  & 25.53 & 27.73 & 7.9\%  & 2.20 \\
Zenodo & 256 & 500  & 1.88 & 7.69  & 75.5\% & 5.80 \\
MC~Maze & 1024 & 500 & 25.53 & 47.45 & 46.2\% & 21.93 \\
Zenodo & 256 & 5000 & 1.88 & 59.89 & 96.9\% & 58.01 \\
MC~Maze & 1024 & 5000 & 25.53 & 244.67 & 89.6\% & 219.15 \\
\bottomrule
\end{tabular}}
\end{table}

\begin{table}[!t]
\centering
\caption{Peak Memory (MB) vs.\ Hidden-Layer Width $N$}
\label{tab:memory_hidden}
\resizebox{\columnwidth}{!}{%
\begin{tabular}{c c c c c c}
\toprule
\textbf{Hidden} & \textbf{Online} & \textbf{BPTT} & \textbf{Dynamic} & \textbf{Reduction} & \textbf{Savings} \\
\midrule
32   & 0.07  & 2.27  & 2.20  & 96.9\% & 2.20  \\
128  & 0.57  & 6.80  & 6.23  & 91.7\% & 6.23  \\
256  & 1.88  & 13.49 & 11.60 & 86.0\% & 11.60 \\
512  & 6.76  & 29.11 & 22.34 & 76.8\% & 22.35 \\
1024 & 25.53 & 69.37 & 43.83 & 63.2\% & 43.84 \\
2048 & 99.05 & 185.87 & 86.80 & 46.7\% & 86.82 \\
4096 & 390.11 & 562.89 & 172.73 & 30.7\% & 172.78 \\
\bottomrule
\end{tabular}}
\end{table}

\section{Hyperparameters and Bin-Width Sensitivity}
\label{app:hyperparams}

Table~\ref{tab:sensitivity} reports hyperparameter sensitivity on Zenodo across five sessions. All Welch $t$-tests against the baseline configuration yield $p > 0.49$, indicating that none of the tested variations produce statistically distinguishable performance. The trace time constants ($\tau_{\mathrm{fast}}$, $\tau_{\mathrm{slow}}$), mixing coefficient ($\alpha_{\mathrm{mix}}$), and consolidation window ($K$) each show spread below 2\%, confirming that the learning rule is robust to moderate perturbations. The learning rates exhibit higher sensitivity (10.6\% spread), consistent with the general observation that step-size tuning matters more than architectural hyperparameters in online learning.

Table~\ref{tab:binwidth} reports bin-width sensitivity on Zenodo. Performance degrades monotonically as bin width increases beyond 50\,ms: at 256\,ms bins, correlations collapse to near zero ($R{=}{-}0.018$ on $X$), indicating that fine temporal resolution is necessary for the learning rule to exploit spike timing structure. The 32\,ms and 50\,ms bins yield comparable performance, suggesting diminishing returns below 50\,ms for this dataset. Table~\ref{tab:hyperparams} provides the complete decoder configurations used across all experiments.

\begin{table}[!t]
\centering
\caption{Hyperparameter Sensitivity on Zenodo (Five Sessions). All $p > 0.49$ (Welch's $t$-tests vs.\ baseline).}
\label{tab:sensitivity}
\resizebox{\columnwidth}{!}{%
\begin{tabular}{l l c c}
\toprule
\textbf{Parameter} & \textbf{Values tested} & \textbf{Baseline} & \textbf{Spread} \\
\midrule
$\tau_{\text{fast}}$ (ms) & 60, 120, 180 & 120 & 1.4\% \\
$\tau_{\text{slow}}$ (ms) & 400, 700, 1000 & 700 & 1.4\% \\
$\alpha_{\text{mix}}$ & 0.1, 0.3, 0.5, 0.7, 0.9 & 0.5 & 1.8\% \\
fast\_lr & $10^{-4}$ to $5{\times}10^{-3}$ & $2{\times}10^{-3}$ & 10.6\% \\
slow\_lr & $10^{-5}$ to $5{\times}10^{-4}$ & $2{\times}10^{-4}$ & 10.6\% \\
window $K$ & 10, 25, 50, 100, 200 & 50 & 1.2\% \\
\bottomrule
\end{tabular}}

\vspace{1.5em}

\caption{Zenodo Bin-Width Sensitivity (Five Sessions)}
\label{tab:binwidth}
\begin{tabular}{c c c}
\toprule
\textbf{Bin (s)} & \textbf{X corr} & \textbf{Y corr} \\
\midrule
0.032 & $0.637 \pm 0.126$ & $0.764 \pm 0.051$ \\
0.050 & $0.611 \pm 0.107$ & $0.727 \pm 0.090$ \\
0.064 & $0.568 \pm 0.133$ & $0.671 \pm 0.144$ \\
0.100 & $0.398 \pm 0.167$ & $0.544 \pm 0.198$ \\
0.128 & $0.424 \pm 0.135$ & $0.412 \pm 0.119$ \\
0.200 & $0.191 \pm 0.066$ & $0.320 \pm 0.210$ \\
0.256 & $-0.018 \pm 0.174$ & $0.154 \pm 0.171$ \\
\bottomrule
\end{tabular}
\end{table}

\begin{table}[!t]
\centering
\caption{SNN Decoder Hyperparameters}
\label{tab:hyperparams}
\resizebox{\columnwidth}{!}{%
\begin{tabular}{l c c c c}
\toprule
\textbf{Parameter} & \textbf{MC Batch} & \textbf{MC True} & \textbf{Zen.\ Batch} & \textbf{Zen.\ True} \\
\midrule
\multicolumn{5}{l}{\textit{Architecture \& Data}} \\
Hidden layers & 1024, 512 & 1024, 512 & 256, 128 & 256, 128 \\
Bin width & 100\,ms & 100\,ms & 50\,ms & 50\,ms \\
Kinematic lag & 80\,ms & 80\,ms & 0\,ms & 0\,ms \\
Input & Raw count & Raw count & Raw count & Raw count \\
\midrule
\multicolumn{5}{l}{\textit{Neuron Parameters}} \\
LIF threshold & 1.0 & 1.0 & 1.0 & 1.0 \\
LIF $\beta$ (hid / out) & 0.7 / 0.5 & 0.7 / 0.5 & 0.7 / 0.5 & 0.7 / 0.5 \\
Surrogate grad. & Fast sigm. & Fast sigm. & Fast sigm. & Fast sigm. \\
Weight cap $c_\ell$ & 6.0 & 6.0 & 6.0 & 6.0 \\
\midrule
\multicolumn{5}{l}{\textit{Learning Parameters}} \\
Fast LR & $1{\times}10^{-5}$ & $1{\times}10^{-4}$ & $3{\times}10^{-3}$ & $2{\times}10^{-3}$ \\
Slow LR & $1{\times}10^{-6}$ & $1{\times}10^{-5}$ & $1{\times}10^{-3}$ & $2{\times}10^{-4}$ \\
$\tau_{\text{fast}}$/$\tau_{\text{slow}}$ & 120/700\,ms & 120/700\,ms & 120/700\,ms & 120/700\,ms \\
$\alpha_{\text{mix}}$ & 0.5 & 0.8 & 0.5 & 0.8 \\
Window $K$ & 50 & 200 & 50 & 50 \\
Momentum $\mu$ & 0.9 & 0.9 & 0.9 & 0.9 \\
Weight decay & $10^{-5}$ & $10^{-5}$ & $10^{-5}$ & $10^{-5}$ \\
\midrule
\multicolumn{5}{l}{\textit{Training}} \\
Epochs & 20 & 3 & 20 & 20 \\
Batch size & 32 & 1 & 32 & 1 \\
Patience & 10 & 8 & 10 & 20 \\
\bottomrule
\end{tabular}}
\end{table}

\begin{table}[!t]
\centering
\caption{Baseline Decoder Hyperparameters}
\label{tab:baselines_hyperparams}
\resizebox{\columnwidth}{!}{%
\begin{tabular}{l c c}
\toprule
\textbf{Parameter} & \textbf{LSTM (both datasets)} & \textbf{KF (both datasets)} \\
\midrule
Hidden layers & 256, 256 & N/A \\
Input & Rate, z-scored & Rate, z-scored \\
Optimizer & Adam ($1{\times}10^{-3}$) & Analytical \\
Weight decay & $10^{-5}$ & -- \\
Epochs & 50 & Analytical \\
Batch size & 64 & N/A \\
Patience & 10 & N/A \\
\bottomrule
\end{tabular}}
\end{table}

\end{document}